\documentstyle[12pt]{article}
\input{amssym.def}

\makeatletter
\@addtoreset{equation}{section}

\makeatother
\begin{document}

\def\g{\frak{g}}
\def\ga{\hat{\g}}
\newtheorem{Lemma}{Lemma}[section]
\newtheorem{Theorem}{Theorem}[section]
\newtheorem{Example}{Example}[section]
\newtheorem{Definition}{Definition}[section]
\newtheorem{Remark}{Remark}[section]
\newtheorem{Proposition}{Proposition}[section]
\def\h{\frak {h}}
\def\ha{\hat{\h}}
\def\n{\frak {n}}
\def\na{\hat{\n}}
\def\nb{\bar{\n}}
\def\C{\Bbb {C}}
\def\N{\Bbb {N}}
\def\Z{\Bbb {Z}}
\newcommand{\be}{\begin{equation}}
\newcommand{\ee}{\end{equation}}

\setcounter{section}{-1}
\hfill hep-th/9412054
\begin{center}\LARGE Combinatorial  constructions of modules for
infinite-dimensional Lie algebras, I. Principal
subspace.\\\vspace{5mm} {\Large Galin Georgiev}\\ {\small Department
of Mathematics, Rutgers University, New Brunswick, NJ 08903\footnote
{georgiev@math.rutgers.edu; Math. Subj. Classific. 17, 82, 05}}
\end{center}\vspace{8mm}
\begin{abstract}\begin{em}
   This is the first of a series of papers studying combinatorial
(with no ``subtractions'') bases and characters of standard modules
for affine Lie algebras, as well as various subspaces and ``coset
spaces'' of these modules.

   In part I we consider certain standard modules for the affine Lie
algebra $\ga,\;\g := sl(n+1,\C),\;n\geq 1,$ at any positive integral
level $k$ and construct bases for their principal subspaces
(introduced and studied recently by Feigin and Stoyanovsky [FS]). The
bases are given in terms of partitions: a color $i,\;1\leq i \leq n,$
and a charge $s,\; 1\leq s \leq k,$ are assigned to each part of a
partition, so that the parts of the same color and charge comply with
certain difference conditions.  The parts represent ``Fourier
coefficients'' of vertex operators and can be interpreted as
``quasi-particles'' enjoying (two-particle) statistical interaction
related to the Cartan matrix of $\g.$ In the particular case of vacuum
modules, the character formula associated with our basis is the one
announced in [FS]. New combinatorial characters are proposed for the
whole standard vacuum $\ga$-modules at level one.
\end{em}\end{abstract}

\section{Introduction}
\subsection{}

   This paper was meant to be a higher-rank generalization of the
seminal work of Lepowsky and Primc [LP2] where they built vertex
operator combinatorial bases for what came to be called {\em $\ga
\supset \ha$ coset subspaces} of standard (integrable highest weight)
$\ga$-modules, $\h$ being the Cartan subalgebra of $\g = sl(2,\C).$
The paper [LP2] was part of the ${\cal Z}$-algebra program originated
in [LW1-3], [LP1-2] for studying affine algebras through vertex
operators centralizing their Heisenberg subalgebras.  Recall that the
structure of the $\ga
\supset \ha$ coset subspaces  is essentially
encoded in their quotient spaces of coinvariants with respect to the
action of a certain abelian group. The latter are known to the
physicists as {\em parafermionic spaces} because they turned out to be
the underlying spaces of the so-called parafermionic conformal field
theories of Zamolodchikov and Fateev [ZF1] (with Virasoro algebra
central charge $c = \frac{2k-2}{k+2},\; k
\in \Z,\;k >1;$ we note that the original twisted vertex operator
construction [LW1] found a conformal-field-theoretical understanding
in [ZF2]). The original ${\cal Z}$-algebra program was later
generalized and elaborated both by mathematicians [C], [Hu], [Ma],
[MP1], [MP2], [Mi1-4], [P1-2], and physicists -- cf. [G] where the
string functions and parafermionic field theories associated with more
general affine Lie algebras were studied. Variations of the celebrated
Rogers-Ramanujan identities and their generalizations -- the
Gordon-Andrews-Bressoud identities [Go], [A] -- were first interpreted
in representation-theoretical terms by Lepowsky and Wilson [LW1-3] and
were ubiquitous in the subsequent works. Similar in spirit
construction -- in terms of the ''difference two'' condition,
discovered in representation-theoretical context in [LW1] -- was later
proposed for the Virasoro algebra minimal models ${\cal
M}(2,2n+1),\;n>1,$ [FNO], [FF].  Despite all the progress, we were
still far away from a simple and conceptual vertex operator
(combinatorial) construction of the higher-level standard modules for
higher-rank affine Lie algebras.  A good measure of this deficiency
was the lack of nice and general enough combinatorial character
formulas.

   In the present paper and its follow-ups we shall venture to fill in
part of this gap. Working in the setting of Vertex Operator Algebra
Theory [FLM], we follow the tradition and build vertex operator
combinatorial bases, i.e., each basis vector is a finite product of
vertex operators acting on a highest weight vector (the highest
weights in consideration here are specified at the beginning of
Section 5).  Although we work with $\g = sl(n+1, \C),$ our
construction has a straightforward generalization for any
finite-dimensional simple Lie algebra $\g$ of type A-D-E. Major
technical tool is the homogeneous vertex operator construction of
level one standard modules [FK], [S].  The most delicate part of the
proofs -- the independence arguments -- employ the intertwining vertex
operators (in the sense of [FHL]) constructed in [DL], which
interchange different modules (but a general idea of J. Lepowsky's, to
build bases from the intertwining vertex operators themselves, is yet
to become a reality; for the level one case, cf.  Proposition \ref{Pro
0.2} below and for the rank one case, cf.  [BLS1-3]).

    We would like to point out that constructing a basis for $\ga
\supset \ha$ coset subspaces (or, equivalently, for the parafermionic spaces)
disentangles also the structure of a variety of other important
representations. For example, the standard module itself is well known
[LW1] to be a tensor product of its $\ga \supset \ha$ coset subspace
and a bosonic Fock space representation of a Heisenberg algebra
associated with $\ha.$ More original application (to be discussed in
[GeIII]) of structural results for parafermionic spaces is through
''nested'' coset subspaces of type $\left(\ga^{(1)} \supset
\ha^{(1)}\right)
\supset \left(\ga^{(2)}
\supset \ha^{(2)}\right)$ for some pair $\g^{(1)} \supset \g^{(2)}.$
If one picks for example $\g^{(2)} := sl(n,\C)$ which is naturally
embedded in $\g^{(1)} := sl(n+1,\C),$ one obtains the Virasoro algebra
unitary modules [FQS] with central charge $c = 1 -
\frac{6}{(n+2)(n+3)},\;n \in \Z,\; n > 1$ (the underlying spaces of
the minimal conformal field theories of Belavin, Polyakov and
Zamolodchikov [BPZ]; as Alexander Zamolodchikov pointed to us, a
similar in spirit coset construction had been largely used by early
conformal field theorists prior to  the ground-breaking work [GKO]).

   For reasons not fully understood, the character formula associated
with the Lepowsky-Primc basis was vastly generalized in different
directions on the physics side through a sweeping series of
fascinating conjectures (inspired mainly by Thermodynamic Bethe Ansatz
techniques): cf. [T], [NRT], [KNS], [KM], [KKMM1], [KKMM2], [DKKMM1],
[DKKMM2], [KMM], [WP], [BM] to name some few.  The announced proofs
[Ber], [Kir], [FQ], [BM], [FW] of some of these conjectures, do not
even attempt to reveal the underlying vertex operator combinatorial
bases (very inspiring exception is the spinon construction of basic
$\widehat{sl}(2,\C)$-modules related to the so-called Haldane-Shastry
spin chain [BPS], [BLS1-3]; see Proposition
\ref{Pro 0.2} at the end of this Introduction where we state the
higher-rank generalization of the spinon character formula for level
one).  Not surprisingly, the true precursor of all these new
characters seems to be another remarkable subspace (of the standard
modules for affine Lie algebras), introduced in a recent announcement
of Feigin and Stoyanovsky [FS]: The so-called {\em principal
subspace}, whose dual space has been beautifully described in [FS] in
terms of symmetric polynomial forms vanishing on certain hyperplanes.
These subspaces are generated by the affinization of the nilpotent
subalgebra of $\g$ consisting of (strictly) upper-triangular matrices
when $\g = sl(n+1,\C)$ and have obvious generalizations for any simple
finite-dimensional Lie algebra $\g.$ It turned out to be more
conceptual and technically easier to work with vertex operators (the
products of whose coefficients generate our bases) in the setting of
principal subspaces. More importantly for our initial commitments,
establishing a vertex operator basis for a principal subspaces implies
the existence of a similar basis for the corresponding parafermionic
space (this issue will be addressed in part II [GeII]; Feigin and
Stoyanovsky themselves were well aware of this close relationship
between their considerations and the approach of Lepowsky, Wilson and
Primc [LW1-3], [LP1-2]).  Furthermore, representing a given standard
module as a direct limit of ''twisted'' (hit by special inner
automorphisms) principal subspaces, one can hope to obtain a
combinatorial basis for the whole standard module in terms of
''semiinfinite monomials'' (see [FS] for the case $\g = sl(2,\C);$ this
approach is similar in spirit to [LP2] and, not surprisingly, the
corresponding character formulas are the same). Actually, there is
more natural, in our opinion, way to build the whole standard module
in terms of {\em finite} monomials, starting from the principal
subspace: One simply has to throw in the game the negative simple
roots and keep playing by the same rules. The corresponding character
formula for the level one case is given in Proposition \ref{Pro 0.1}
at the end of this Introduction. The higher levels are treated in
[GeII].

\subsection{} We proceed with a  demonstration of our bases in the two
simplest  examples beyond $\widehat{sl}(2,\C)$: the principal
subspaces of the vacuum standard $\widehat{sl}(3,\C)$-modules at
levels one and two. They carry to a large extent  the spirit of the
general case and will ease the impetuous reader.

   Fix a triangular decomposition $sl(3,\C) =: \g = \n_{-} \oplus \h
\oplus
\n_{+}$ and a corresponding Chevalley basis of $\g:$
$$\{x_{-\alpha_{1}},x_{-\alpha_{2}},
x_{-\alpha_{1}-\alpha_{2}}\}\;\bigsqcup\; \{h_{\alpha_{1}},
h_{\alpha_{2}}\}\; \bigsqcup\; \{x_{\alpha_{1}},x_{\alpha_{2}},
x_{\alpha_{1}+\alpha_{2}}\},$$ where $\alpha_{1}, \alpha_{2}$ are the
simple (positive) roots. Let $\ga = \g \otimes \C [t^{-1},t] \oplus \C
c$ be the corresponding (untwisted) affine Lie algebra (cf. [K]) and
set $g(m) := g \otimes t^{m},\; D:= -td/dt, \;g \in \g,\; m \in \Z.$
Let $v(\hat{\Lambda}_{0})$ be the highest weight vector of the vacuum
standard $\ga$-module $L(\hat{\Lambda}_{0})$ at level one (the
eigenvalue of the central charge $c$ is called level). Motivated by
[DKKMM1] and [FS], we shall refer to $x_{\alpha_{i}}(m), \; i =1,2,$ as
{\em quasi-particle of charge $1,$ color $i$ and energy $-m.$} The
vacuum principal subspace at level one $W( \hat{\Lambda}_{0}):=
U(\n_{+} \otimes \C [t^{-1},t])\cdot v(\hat{\Lambda}_{0})$ (where
$U(\cdot)$ denotes universal enveloping algebra), has a basis
generated by the following color ordered quasi-particle monomials
acting on $v(\hat{\Lambda}_{0}):$
\be \frak{B}_{W(\hat{\Lambda}_{0})} = \bigsqcup_{r_{2},r_{1} \geq 0}
\left\{ x_{\alpha_{2}}(m_{r_{2},2})\cdots
x_{\alpha_{2}}(m_{1,2})x_{\alpha_{1}}(m_{r_{1},1}) \cdots
x_{\alpha_{1}}(m_{1,1})
\left | \begin{array} {l}\\ \\  \end{array}\right.\right.\label{0.1}\ee
$$\left|\begin{array}{l} m_{p,i} \in
r_{i-1} - 1 - \N\;\;\mbox{for}\;\; 1 \leq p \leq r_{i} ;\\m_{p+1,i} \leq
m_{p,i} -2 \;\;\mbox{for}\;\; 1\leq p <  r_{i};\; i =1,2
\end{array}\right\}$$
where $r_{0} := 0$ and $\N := \{0,1,2,\ldots\}.$ For an explicit list
of some basis elements with low enegies, see Example 4.1, Section 4
and the corresponding Table 1 in the Appendix. The quasi-particle
monomial basis for $\g = sl(n+1,\C)$ and generic fundamental (i.e.
level one) highest weight $\hat{\Lambda}_{j},\; 0\leq j \leq n,$ is
given in Definition \ref{Def 3.1} and formula (\ref{3.101}).

   In physicists' terms, the above basis can be described in a very
simple and natural way, reminiscent of the ''functional'' description
of the restricted dual of the principal space in [FS]: Consider the
Fock space of two different (of color $1$ or $2$) free bosonic
quasi-particles with a single quasi-particle energy spectrum
consisting of all the integers greater than or equal to the charge (= 1) of
the quasi-particle. Take a hamiltonian ${\cal H}$ which is simply a
sum of a single-particle term ${\cal H}_{1}$ and a two-particle
interaction term ${\cal H}_{2}$ (this interaction should be thought of
as a statistical interaction in the sense of Haldane [H]). The
single-particle energy is of course the sum of the single-particle
energies of all the quasi-particles (in a given state) and the
two-particle energy is a sum of the interaction energies of all the
pairs of quasi-particles (in a given state).  The energy of
interaction between a quasi-particle of color $l$ and another
quasi-particle of color $m$ is $A_{lm},$ where $(A_{lm}) :=
\left(\begin{array}{cc} \scriptstyle{2}&
\scriptstyle{-1}\\ \scriptstyle{-1} & \scriptstyle
{2}\end{array}\right)$ is the Cartan matrix of $\g $ (the inner
products of simple roots are inherently encoded in both the
spanning and independence part of our proof; cf. Section 4).  Since
the character of the Fock space of noninteracting quasi-particles is
\be \mbox{\rm Tr}\,q^{{\cal H}_{1}} \left| \begin{array}[t]{l} \\
\mbox{Fock} \end{array} \right.  = \frac{1}{(q)_{\infty}^{2}} =
\sum_{r_{1}\geq 0} \frac{q^{r_{1}}}{(q)_{r_{1}}}\,\sum_{r_{2}\geq 0}
\frac{q^{r_{2}}}{(q)_{r_{2}}}= \label{0.2}\ee
$$= \sum_{r_{1}, r_{2} \geq 0} \frac{q^{r_{1} +
r_{2}}}{(q)_{r_{1}}(q)_{r_{2}}},$$ where $(q)_{r} :=
(1-q)(1-q^{2})\cdots (1-q^{r}),\; (q)_{0} := 1,$ the full
$q$-character of $W(\hat{\Lambda}_{0})$ will be a ''deformation'' of
this expression with an interaction term (the binomial coefficient
$\left(\begin{array}{c}\scriptstyle{ r_{i}}\\ \scriptstyle{ 2}
\end{array}\right)$ counting all the pairs of color $i):$
\be  \mbox{\rm Tr}\,q^{D} \left| \begin{array}[t]{l} \\
W(\hat{\Lambda}_{0}) \end{array} \right. = \mbox{\rm Tr}\,q^{{\cal H}_{1}+
{\cal H}_{2}} \left| \begin{array}[t]{l} \\
\mbox{Fock} \end{array} \right.  =  \sum_{r_{1}, r_{2} \geq 0} \frac{q^{r_{1} +
r_{2} + 2\left(\begin{array}{c}\scriptstyle{ r_{1}}\\\scriptstyle{ 2}
\end{array}\right) + 2\left(\begin{array}{c}\scriptstyle{ r_{2}}\\
\scriptstyle{ 2}
\end{array}\right) - r_{1}r_{2}
}}{(q)_{r_{1}}(q)_{r_{2}}} = \label{0.101}\ee $$= \sum_{r_{1}, r_{2}
\geq 0} \frac{q^{r_{1}^{2} + r_{2}^{2} -
r_{1}r_{2}}}{(q)_{r_{1}}(q)_{r_{2}}} = \sum_{r_{1}, r_{2}
\geq 0} \frac{q^{\frac{1}{2} \sum_{l,m = 1}^{2} A_{lm}r_{l}
r_{m}}}{(q)_{r_{1}}(q)_{r_{2}}}.$$
The last expression is the Feigin-Stoyanovsky formula [FS] (for
generic rank and fundamental highest weight, see (\ref{3.14}) here).
At the end of the Introduction we state  the analog of this
formula for the {\em whole} vacuum basic  module $L(\hat{\Lambda}_{0}).$

   We proceed with the basis for the level two vacuum principal
subspace ($\g = sl(3,\C)).$ Here comes the surprise: rather
unexpectedly, we shall give up the traditional approach to build
module basis using a basis of the Lie algebra itself (i.e.,
quasi-particles of charge one).  Instead, the building blocks of our
basis will be certain -- very natural {}from the point of view of
Conformal Field Theory (and for that matter, Vertex Operator Algebra
Theory [FLM]) -- infinite linear combination of
charge-one-quasi-particle monomials [LP2], [ZF1]. The use of these
higher-charge (see below) vertex operators changes dramatically the
structure of the basis: even in the special case $\g = sl(2,\C), $ we
end up with construction very different from [LP2] for example (cf.
[GeII] for more detailed discussion). We believe that only this
broader perspective can save the elegance and simplicity of the
picture in the presence of more than one color (i.e., rank$\,\g > 1$)
at higher levels. The linear combinations in question are actually
truncated (i.e., finite) when acting on highest weight modules and can
be thought of as {\em quasi-particles of charge $>1$}. For example, a
quasi-particle of charge $2,$ color $i$ and energy $-m$ is defined as
follows
\be x_{2\alpha_{i}}(m) := \sum_{\begin{array}{c}{\scriptstyle m_{2},
m_{1} \in \Z }\\{\scriptstyle m_{2} + m_{1} = m}
\end{array}}x_{\alpha_{i}}(m_{2})x_{\alpha_{i}}(m_{1})\label{0.3}\ee
(for a general definition of quasi-particles of arbitrary charge, see
Section 3). Note that a quasi-particle of charge $2$ ''confines'' two
quasi-particles of charge $1$ in such a way that only the {\em total}
energy can be read off (the individual energies are not
''measurable''). In terms of the vertex operator (bosonic current)
\be X_{\alpha_{i}}(z) := Y(x_{\alpha_{i}}(-1) \cdot
v(2\hat{\Lambda}_{0}) ,z) = \sum_{m\in \Z} x_{\alpha_{i}}(m)
z^{-m-1}\label{0.4}\ee ($z$ is simply a formal variable and
$v(2\hat{\Lambda}_{0})$ is the highest weight vector of the level two
vacuum standard module $L(2\hat{\Lambda}_{0})),$ one has
\be X_{2\alpha_{i}}(z) :=  X_{\alpha_{i}}(z)X_{\alpha_{i}}(z) =
Y(x_{\alpha_{i}}(-1)^{2}\cdot v(2\hat{\Lambda}_{0}), z) = \sum_{m\in
\Z} x_{2\alpha_{i}}(m) z^{-m-2}.\label{0.6}\ee This is a standard
recipe in Conformal Field Theory for obtaining currents of higher
charge (the usual ''normal ordering'' of the product
$X_{\alpha_{i}}(z)X_{\alpha_{i}}(z)$ is not needed here because of the
commutativity of the two factors).

   Let us now come back to   the vacuum principal subspace at level two $W(2
\hat{\Lambda}_{0}):= U(\n_{+} \otimes \C
[t^{-1},t])\cdot v(2\hat{\Lambda}_{0}).$ Our basis will be generated
by (charge and color ordered) quasi-particle monomials acting on
$v(2\hat{\Lambda}_{0}),$ with $p_{i}^{(2)} := r_{i}^{(2)}$
quasi-particles of charge $2$ and color $i,\; i =1,2,$ and
$p_{i}^{(1)} := r_{i}^{(1)} -r_{i}^{(2)}$ quasi-particles of charge
$1$ and color $i,$ for some $r_{i}^{(1)} \geq r_{i}^{(2)}\geq 0$
(i.e., a total charge of $p_{i}^{(1)} + 2p_{i}^{(2)} = r_{i}^{(1)} +
r_{i}^{(2)}).$ The conditions they must satisfy are given below (the
top two lines on the right-hand side of the delimiter $|$ concern the
quasi-particles of charge $2$; the bottom two lines concern the
quasi-particles of charge $1$):
\be \frak{B}_{W(2\hat{\Lambda}_{0})}= \bigsqcup_{\begin{array}{c}{\scriptstyle
r_{2}^{(1)} \geq r_{2}^{(2)} \geq 0}\\{\scriptstyle r_{1}^{(1)} \geq
r_{1}^{(2)} \geq 0}\end{array}} \label{0.7}\ee
$$\left\{x_{\alpha_{2}}(m_{r_{2}^{(1)},2})\cdots
x_{\alpha_{2}}(m_{r_{2}^{(2)}+1,2}) x_{2 \alpha_{2}}(
m_{r_{2}^{(2)},2}) \cdots x_{2\alpha_{2}}(m_{1,2})\cdot
\begin{array}{c}\\ \\
\end{array}\right.$$
$$\left. \cdot x_{\alpha_{1}}(m_{r_{1}^{(1)},1})\cdots
x_{\alpha_{1}}(m_{r_{1}^{(2)}+1,1}) x_{2\alpha_{1}}(
m_{r_{1}^{(2)},1}) \cdots x_{2\alpha_{1}}(m_{1,1})\begin{array}{c}\\
\\
\end{array}\right|$$
$$\left| \begin{array}{l} m_{p,i} \in r_{i-1}^{(1)} + r_{i-1}^{(2)} -
2- \N \;\;\mbox{for}\;\; 1 \leq p \leq r_{i}^{(2)};\\ m_{p+1,i} \leq
m_{p,i} -4\;\;\mbox{for}\;\;1 \leq p < r_{i}^{(2)};\;i =1,2;\\
\\m_{p,i} \in r_{i-1}^{(1)} - 2r_{i}^{(2)} - 1 -
\N\;\;\mbox{for}\;\;r_{i}^{(2)} < p \leq r_{i}^{(1)} ;\\ m_{p+1,i}
\leq m_{p,i} - 2\;\;\mbox{for}\;\; r_{i}^{(2)} < p < r_{i}^{(1)};\;i
=1,2\end{array}\right\},$$ where $r_{0}^{(1)} = r_{0}^{(2)} := 0.$ For
an explicit list of some basis elements with low enegies, see Example
5.1, Section 5 and the corresponding Table 2 in the Appendix. (The
quasi-particle monomial basis for $\g = sl(n+1,\C)$ and level $k$
highest weight $\hat{\Lambda} = k_{0}\hat{\Lambda}_{0} +
k_{j}\hat{\Lambda}_{j},\; k_{0} + k_{j} = k,\; 1\leq j \leq n,$ is
given in Definition \ref{Def 4.1} and formula (\ref{4.12}).)

   Although the initial conditions above may look strange at first
sight, they can be read off in a remarkably simple way (a
straightforward generalization of the level one picture): Consider the
Fock space of four different (of color $1$ or $2$ and charge $1$ or
$2$) free bosonic quasi-particles with single quasi-particle energy
spectrum consisting of all the integers greater or equal to the charge
of the quasi-particle.  The hamiltonian consists again only of a
single-particle term ${\cal H}_{1}$ and a two-particle interaction
term ${\cal H}_{2}.$ The single-particle energy is a sum of the
single-particle energies of all the quasi-particles (in a given state)
and the two-particle energy is a sum of the interaction energies of
all the pairs of quasi-particles (in a given state). The energy of
interaction between a quasi-particle of charge $s$ and color $l$ and
another quasi-particle of charge $t$ and color $m$ is
$A_{lm}\,\mbox{min}\{s,t\}.$ Similarly to the level one case, the
corresponding $q$-character is
\be  \mbox{\rm Tr}\,q^{D} \left| \begin{array}[t]{l} \\
W(2\hat{\Lambda}_{0}) \end{array} \right. = \mbox{\rm Tr}\,q^{{\cal
H}_{1}+ {\cal H}_{2}} \left|
\begin{array}[t]{l} \\ \mbox{Fock} \end{array} \right. =
  \sum_{\begin{array}{c}{\scriptstyle p_{1}^{(1)}, \;p_{1}^{(2)} \geq
0}\\{\scriptstyle p_{2}^{(1)}, \;p_{2}^{(2)} \geq 0 }\end{array}}
\frac{q^{\frac{1}{2}\sum_{l,m =
1,2}^{s,t = 1,2}\;A_{lm} B^{st}
p_{l}^{(s)}p_{m}^{(t)}}}{(q)_{p_{1}^{(1)}} (q)_{p_{1}^{(2)}}
(q)_{p_{2}^{(1)}}(q)_{p_{2}^{(2)}} } ,\label{0.8}\ee where $B^{st} :=
\mbox{min}\{s,t\},$ $1\leq s,t \leq 2$ (cf. formula (\ref{4.27}) for
the general case). This is the  character announced by
Feigin and Stoyanovsky  [FS].

\subsection{}A few words should be said about the continuation of
part I and its easily-conceivable generalizations, as well as some
related open problems.

   As already mentioned, part II [GeII] employs the given here bases
and constructs similar quasi-particle monomial bases for the
parafermionic spaces in standard modules ($ \g = sl(n+1,\C)$ and
highest weights like the ones considered here). In the particular case
of the vacuum module, the associated character formula is the
$\widehat{sl}(n+1,\C)$-case of Kuniba-Nakanishi-Suzuki conjecture
[KNS] (for the vacuum module, an independent proof using dilogarithms
was recently announced in [Kir]).

   In [GeIII] we shall ''factorize'' these characters in order to
obtain combinatorial characters for the nested coset $\left(\ga^{(1)}
\supset
\ha^{(1)}\right) \supset \left(\ga^{(2)} \supset \ha^{(2)}\right)$
subspace of two $\ga \supset \ha$ coset subspaces of level two
standard modules, $sl(n+1,\C) = \g^{(1)} \supset \g^{(2)} = sl(n,\C).$
With the natural sructure of Virasoro algebra modules, these nested
coset spaces are exactly the unitary modules of central charge $c = 1
- \frac{6}{(n+2)(n+3)},\;n \in \Z,\; n > 1,$ (in other words, we use a
coset realization different from the classical [GKO]).
The obtained combinatorial characters are the ''fermionic'' characters
conjectured by Kedem, Klassen, McCoy and Melzer [KKMM2] (cf. also
[M1]; for a large subclass of the modules in question, these character
formulas were proven by very different methods in [Ber]; general
minimal models are treated in [BM]). In order to
cover all the Virasoro algebra modules from this series (not only the
vacuum ones!), one needs to go beyond the {\em vacuum} module
characters of Feigin-Stoyanovsky and Kuniba-Nakanishi-Suzuki and
consider more general dominant integral highest weights.

   Coming back to the considerations in part I, we would like to point
out that they have a straighforward generalization for the untwisted
affinization $\ga$ of any finite-dimensional simple Lie algebra $\g$
of type A-D-E and highest weights of the type considered here. What is
not clear at this point is how to generalize elegantly this
construction for {\em any} dominant integral highest weight? Another
exciting open problem is to construct a basis of colored semi-infinite
quasi-particle monomials for the whole standard module, thus
generalizing the $\widehat{sl}(2,\C)$-considerations of Feigin and
Stoyanovsky [FS] (the generated in this way characters are the same as
the ones obtained from the $\ga \supset \ha$ coset decomposition of
the standard module and the the Kuniba-Nakanishi-Suzuki characters of
the parafermionic spaces).

   We conclude this Introduction with the presentation of two new
combinatorial character formulas for the vacuum basic $\ga$-module
$L(\hat{\Lambda}_{0})$ (in order to be coherent with the rest of the
paper, we shall state the results for $\g = sl(n+1,\C),$ but the
generalization for any A-D-E type algebra is obvious).  The first
formula reflects a basis which is the most natural extension of the
constructed here basis for the principal subspace
$W(\hat{\Lambda}_{0}) \subset L(\hat{\Lambda}_{0}):$ One simply has to
imitate the principal subspace construction, adding vertex operators
corresponding to the negative simple roots and taking into account the
new constraint (cf. [LP2])
$$(z_{2} - z_{1})^{\langle \alpha_{i},
\alpha_{i}\rangle} X_{-\alpha_{i}}(z_{2})
X_{\alpha_{i}}(z_{1})\left|\begin{array}[t]{c}\\\scriptstyle z_{1} =
z_{2}
\end{array}\right. = \mbox{const}$$
for every simple root $\alpha_{i},\; 1\leq i \leq n.$\vspace{5mm}
\begin{Proposition}One has the following $q$-character for  the vacuum
basic (standard, level one) $\ga$-module:
\be  \mbox{\rm Tr}\,q^{D} \left| \begin{array}[t]{l} \\
L(\hat{\Lambda}_{0}) \end{array} \right. = \sum_{r_{\pm 1}, \ldots , r_{\pm
n} \geq 0} \frac{q^{\frac{1}{2}\sum_{l,m =1}^{n} A_{lm} (r_{+l} -
r_{-l})(r_{+m} - r_{-m}) +
\sum_{l=1}^{n}r_{+l}r_{-l}}}{\prod_{l=1}^{n} (q)_{r_{+l}}
(q)_{r_{-l}}} ,\label{0.9}\ee where $(q)_{r} := (1-q)(1-q^{2})\cdots
(1-q^{r}),\; (q)_{0} := 1$ and $(A_{lm})$ is the Cartan matrix of
$\g.$\label{Pro 0.1}\end{Proposition}\vspace{5mm} Note that in the
particular case $\ga = \widehat{sl}(2,\C),$ one can redefine $r_{1} :=
r_{+1},\; r_{2} := r_{-1},$ and thus obtain the character
(\ref{0.101}) $$ \sum_{r_{1}, r_{2} \geq 0} \frac{q^{r_{1}^{2} +
r_{2}^{2} - r_{1}r_{2}}}{(q)_{r_{1}}(q)_{r_{2}}}$$ of the principal
subspace of the vacuum basic $\widehat{sl}(3,\C)$-module.  This
coincidence was observed and explained by Feigin and Stoyanovsky [FS]
(the expression (\ref{0.101}), representing the character of
$L(\hat{\Lambda}_{0})$ for $\ga = \widehat{sl}(2,\C),$ appeared also
in [M2] as a limit of certain finite ''fermionic'' sums). The proof of
the general formula in the above Proposition is analogous to the proof
of the special case [FS]: Using the ''Durfee rectangle'' combinatorial
identity [A]
\be \frac{1}{(q)_{\infty}} := \prod_{l \geq 0} (1- q^{l})^{-1} =
\sum_{\begin{array}[t]{c}\scriptstyle a, b \geq 0\\\scriptstyle a - b
= {\rm const}\end{array} }\frac{q^{ab}}{(q)_{a} (q)_{b}},
\label{0.11}\ee
one immediately checks that (\ref{0.9}) equals the well-known
character expression due to Feingold-Lepowsky [FL] (which inspired  the
homogeneous vertex operator construction [FK], [S])
\be  \mbox{\rm Tr}\,q^{D} \left| \begin{array}[t]{l} \\
L(\hat{\Lambda}_{0}) \end{array} \right. = \frac{1}{(q)_{\infty}^{n}}
\sum _{\beta \in Q}q^{\frac{1}{2}\langle \beta , \beta
\rangle},\label{0.12}\ee
where $Q$ is the root lattice of $\g.$

   The higher level generalization of Proposition \ref{0.1} will be
presented in [GeII] because it requires the respective generalization of
formula (\ref{0.12}).

   Our  second character formula mirrors  another basis
for  the same module, this time   built up from  intertwining
vertex operators (in the sense of [FHL])
corresponding to the fundamental weights and their negatives.\vspace{5mm}
\begin{Proposition}Another expression for the $q$-character of  the vacuum
basic $\ga$-module is
\be  \mbox{\rm Tr}\,q^{D} \left| \begin{array}[t]{l} \\
L(\hat{\Lambda}_{0}) \end{array} \right. = \sum_{\begin{array}[t]{c}
\scriptstyle r_{\pm 1}, \ldots , r_{\pm
n} \geq 0\\ \scriptstyle \sum_{m=1}^{n} A^{(-1)}_{lm}(r_{+m} - r_{-m})
\in \Z\; \forall\;l\end{array}} \frac{q^{\frac{1}{2}\sum_{l,m =1}^{n}
A^{(-1)}_{lm} (r_{+l} - r_{-l})(r_{+m} - r_{-m}) +
\sum_{l=1}^{n}r_{+l}r_{-l}}}{\prod_{l=1}^{n} (q)_{r_{+l}}
(q)_{r_{-l}}} ,\label{0.10}\ee where $(A^{(-1)}_{lm})$ is the inverse
of the Cartan matrix of $\g.$\label{Pro
0.2}\end{Proposition}\vspace{5mm} In the particular case $\ga =
\widehat{sl}(2,\C),$ one can again substitute $r_{1} := r_{+1},\;
r_{2} := r_{-1},$ and thus obtain the Kedem-McCoy-Melzer formula
[KMM], [M2] $$\sum_{\begin{array}[t]{c}\scriptstyle r_{1}, r_{2} \geq
0\\
\scriptstyle r_{1} - r_{2}\; {\rm even} \end{array}}
\frac{q^{ \frac{1}{4}(r_{1} + r_{2})^{2}}}{(q)_{r_{1}}
(q)_{r_{2}}},$$ which was shown to correspond to a spinon basis in
[BPS] (cf. also [BLS1-3]). Note that its higher-rank generalization
(\ref{0.10}) follows again from (\ref{0.11}) and (\ref{0.12}) (this
time, the elements of the root lattice are expressed in terms of
fundamental weights).

   The higher-level generalization of Proposition \ref{0.2} is yet
unknown (one possible approach in the particular case $\g = sl(2,\C)$
was proposed  in [BLS2]).

   The bases underlying the above  expressions will be discussed in
detail elsewhere.

\subsection{}The paper is organized as follows. In Section 1 we
introduce most of our notations and definitions. In Section 2 we
recall the homogeneous vertex operator construction of the basic
(i.e., level one standard) modules. Section 3 introduces the concept
of quasi-particle of any integral positive charge. In Section 4 we
build quasi-particle bases for the principal subspaces of the basic
modules and supply the accompanying character formulas.  Section 5
generalizes this construction and provides quasi-particle bases and
corresponding characters for the principal subspaces at any positive
integral level $k$.  The Appendix contains two tables which illustrate
the examples discussed in this Introduction: Example 4.1 (Section 4)
and Example 5.1 (Section 5).\vspace{5mm}

{\em Acknowledgments} We are very indebted to James Lepowsky for
innumerable suggestions and for corrections and improvements on the
draft of the paper.

   The author was supported by an Excellence Fellowship from the  Rutgers
University Graduate School.

\section {Preliminaries}

\hspace{3ex}We shall use the notation $\Z_{+}$ for the set of positive
integers and $\N$ for $\Z_{+}\, \cup \, \{0\}.$

   Fix $n \in \Z_{+}$ and let $\g := sl(n+1,\C).$ Choose a triangular
decomposition $\g = \n_{-}
\oplus \h \oplus \n_{+}$. Denote by $\Pi :=
\{ \alpha_{1},\ldots ,\alpha_{n} \}$ the set of simple (positive)
roots, with the indices reflecting their locations on the Dynkin
diagram. The notation $\Delta _{+}$ (respectively, $\Delta_{-})$
signifies the set of positive (resp., negative) roots; $\Delta :=
\Delta_{+} \cup
\Delta_{-}$. The highest (positive) root will be denoted by $\theta$ and
let us normalize the nonsingular invariant symmetric bilinear form
$\langle\cdot,\cdot\rangle$ on $\h^{*}\;$ so that $\langle\theta ,
\theta\rangle =2$ (a condition true for any root); then we have a
corresponding form $\langle\cdot,\cdot\rangle$ on $\h.$ Let $\rho$ be
half the sum of the positive roots and $ h = h^{\vee} = n+1$ the
(dual) Coxeter number. For any $\mu \in \h^{*}$ denote by $h_{\mu}
\in \h$ its ``dual'' : $\lambda (h_{\mu} ) = \langle\lambda , \mu
\rangle $ for every
$\lambda \in \h^{*}.$

   We fix for concreteness a Chevalley basis $\, \{ x_{\alpha}
\} _{\alpha \in \Delta} \,\cup\,\{h_{\alpha_{i}}\}_{i=1}^{n}$ of $\g$.
Let $Q :=
\sum_{i=1}^{n} \Z \alpha_{i},$  $P := \sum_{i=1}^{n} \Z \Lambda_{i}$
be the root and weight lattice respectively, where $\Lambda
_{i},\;i=1,\ldots,n,$ are the fundamental weights: $\langle
\Lambda_{i}, \alpha_{j}\rangle  = \delta_{ij},\,i,j = 1,\ldots ,n.$
Set $Q_{+} \subset Q$ (respectively, $Q_{-}) \subset Q$ to be the
semigroup (without $0$) generated by the simple roots $\Pi$
(respectively, by $-
\Pi).$ We denote the group algebras corresponding to $Q$ and $P$ by
$\C [Q] := \mbox{\rm span}_{\C} \{ e^{\beta}
\mid \beta \in Q \}$ and $\C [P] := \mbox{\rm span}_{\C} \{ e^{\lambda}
\mid \lambda \in P \}.$ There exists a central extension
\be 1\;\longrightarrow\;\langle e^{\pi i/(n +
1)^{2}}\rangle\;\longrightarrow\;\hat{P} \;\longrightarrow \;
P\;\longrightarrow \; 1 \label{1.1}\ee (which after restriction
provides a central extension $\hat{Q}$ of $Q),$ by the finite cyclic
group $\langle e^{\pi i/(n + 1)^{2}}\rangle$ of order $2(n + 1)^{2},$
satisfying the following condition: if one chooses a 2-cocycle
\be \varepsilon\,:\;P\,\times\,P \;\longrightarrow\; \langle
e^{\pi i/(n + 1)^{2}}\rangle\label{1.2}\ee
corresponding to the extension,
then one
has
\be \varepsilon (\alpha , \beta)\,\varepsilon (\beta ,
\alpha)^{-1}\,=\,(-1)^{\langle\alpha , \beta
\rangle}\;\;\;\;\mbox{for}\;\;\alpha,\beta \in Q.\label{1.3}\ee
We adopt the notation
\be c(\lambda ,
\mu )\,:= \varepsilon (\lambda , \mu )\,\varepsilon (\mu , \lambda
)^{-1}\;\;\;\;\mbox{for}\;\;\lambda , \mu  \in  P;\label{1.4}\ee
this is the bimultiplicative alternating
commutator map of the central extension (cf. [FLM] Chapter 5 and  [DL]
Chapters 2, 13).

   The affine Lie algebra $\ga $ (of type  $A_{n}^{(1)}$) is the
infinite-dimensional Lie algebra $\ga := \g \otimes \C[t \, ,t^{-1}]
\oplus \C c$ with bracket given by
\be [x \otimes t^{r}, y \otimes t^{s}] = [x \, ,y] \otimes t^{r+s} \, +
\langle x,y\rangle\, r \, \delta_{r+s ,0}\, c ,\label{1.5}\ee
where $x,y \in \g,\; r,s \in \Z$ and $c$ is central. One also needs
the grading
operator $D := -t \, d/dt$. Denote the set of simple (positive) roots
of $\ga$ by $\hat{\Pi} :=
\{ \alpha_{0},\alpha_{1},\ldots ,\alpha_{n} \} \subset (\h \,\oplus\, \C c\,
\oplus\,\C D)^{*}.$ The usual extensions
of $\langle\cdot,\cdot\rangle$ on $\h \,\oplus\, \C c\,
\oplus\,\C D$ and on $(\h \,\oplus\, \C c\,
\oplus\,\C D)^{*}$ will be denoted by the same symbol (we take
$\langle c , D\rangle = -1).$

   We shall often  be working  with the nilpotent subalgebras $\n_{\pm} :=
\mbox{\rm span}_{\C} \{x_{\alpha} \mid \alpha \in \Delta _{\pm} \}
\subset \g$ and the corresponding subalgebra $\nb_{pm} := \n_{pm}\otimes
\C[t \, ,t^{-1}]$ (without $c)$ of $\ga$ (the affinization of
$\n_{\pm}$ is denoted $\na _{\pm} :=  \n_{\pm} \otimes \C[t \, ,t^{-1}]
\oplus \C c \, \subset \ga).$  For the sake of building our basis we
shall need the one-dimensional subalgebras $\n_{\beta} := \C
x_{\beta},
\;\beta \in \Delta_{\pm},$ of $\g$ and the respective abelian
subalgebras $\nb _{\beta} :=
\n_{\beta} \otimes \C[t \, ,t^{-1}]$ of $\ga$  (as opposed to the
affinizations $\na _{\beta} :=  \n_{\beta} \otimes \C[t \, ,t^{-1}]
\oplus \C c $). A crucial ingredient of the vertex operator construction
of basic $\ga$-modules is another affine Lie algebra: the subalgebra
$\ha := \h \otimes \C[t \, ,t^{-1}]
\oplus \C c$ of $\ga.$

   Recall  that for a fixed level $k \in \Z_{+}$ (the scalar by which
$c$ will act on a module), the set of dominant integral weights of
$\ga$  is $\{\sum_{j=0}^{n} k_{j} \hat{\Lambda}_{j} \mid k_{j} \in \N ,\;
\sum_{j=0}^{n}
k_{j} = k \},$ where $\hat{\Lambda}_{i}\,\in \,(\h \,\oplus\, \C c\,
\oplus\,\C D)^{*},\;i=0,\ldots,n,$ are the
fundamental weights of $\ga,$ i.e., $\langle
\hat{\Lambda}_{i},\alpha_{j}\rangle  =
\delta_{ij}$ and $\hat{\Lambda}_{i}(D) = 0,\;i,j = 0,\ldots ,n.$ The
notation $ L(\hat{\Lambda})$ will signify the standard (integrable
irreducible highest weight) $\ga$-module of level $k$ and highest
weight $\hat{\Lambda} =
\sum_{j=0}^{n} k_{j}
\hat{\Lambda}_{j},$ where $ \sum_{j=0}^{n}
k_{j} = k.$

   Let $v(\hat{\Lambda})$ be a highest weight vector of
$L(\hat{\Lambda}).$ Following Feigin-Stoyanovsky [FS], we define the
{\em principal subspace}
\be W(\hat{\Lambda}) := U(\na_{+})\cdot v(\hat{\Lambda}) = U(\nb_{+})\cdot
v(\hat{\Lambda}),\label{1.6}\ee
  where
$U(\cdot )$  always denotes universal
enveloping algebra (similarly, $S(\cdot)$ always denotes a symmetric
algebra).  The principal subspace is defined the same way for
any highest weight module.

   For $k\,\in\,\C^{\times},$ consider the induced $\ha$-module
\be M(k)\,:=\,U(\ha)\,\otimes_{U(\h \otimes \C[t]
\oplus \C c)}\,\C,\label{1.7}\ee
with $\h \otimes \C[t]$ acting trivially
on $\C$ and $c$ acting as $k.$ It is naturally isomorphic as a vector
space to the symmetric algebra $S(\ha^{-}) $, where $\ha^{-} := \h
\otimes t^{-1} \C[t^{-1}]$ (similarly, set  $\ha^{+} := \h
\otimes t \C[t]).$

   For a given $s \in \Z_{+},$ we shall be referring to the
``difference two at distance s'' condition, defined as follows:
suppose we have a sequence of integers, nonincreasing from right to
left,
\be m_{r} \leq \ldots \leq m_{2}\leq m_{1},\label{1.8}\ee
which are indexed
according to their place in the sequence, counted from right to left
(one can think of  such sequence as a partition of the sum of its entries).
The sequence is said to satisfy the ``difference two at distance s''
condition, $s \in \Z_{+},\; s < r,$  if
\be m_{t+s}\leq m_{t} - 2,\;1 \leq t \leq r - s.\label{1.9}\ee

   The following strict linear (lexicographic)  ordering ''$<$'' and
strict partial ordering ''$\prec$'' (called also ''multidimensional''
in [LW2]) will be
largely used in our arguments: For given $r_{n} ,\ldots, r_{1}\in
\Z_{+},$ $ \sum_{i=1}^{n} r_{i} = r,$  consider color-ordered  sequences of
$r_{n}$ integers of ''color'' $n,\; \ldots, $ $r_{1}$ integers of
color $1$:
\be m_{r} \leq \cdots \leq m_{\sum_{i=1}^{n-1}r_{i} +1},
m_{\sum_{i=1}^{n-1}r_{i}} \leq \cdots \leq m_{r_{1}+1}, m_{r_{1}} \leq
\ldots \leq m_{1},\label{1.10}\ee such that only the entries of the
same color are nonincreasing from right to left. For two such
sequences, we write
\be (m_{r},\; \ldots ,\; m_{2},\; m_{1}) < (m'_{r},\; \ldots ,\; m'_{2},\;
m'_{1})\label{1.11}\ee
if there exists $ s \in \Z_{+},\; 1 \leq s \leq r,$ such
that $ m_{1} = m'_{1},\;m_{2} = m'_{2}\;,\ldots, m_{s-1}
= m'_{s-1}\;$ and $m_{s} < m'_{s}.$ On the other hand, we write
\be (m_{r},\; \ldots ,\; m_{2},\; m_{1}) \prec  (m'_{r},\; \ldots ,\; m'_{2},\;
m'_{1})\label{1.12}\ee
if for every $s,\; 1\leq s \leq r,$ one has $m_{s} + \cdots + m_{2} + m_{1}
\leq  m'_{s} + \cdots + m'_{2} + m'_{1}$ and for at least one such $s,$ this
inequality is strict.

   It is easy to see that $a \prec b$ implies $a < b$ but not vice versa.

   We shall also encounter  more general situations when an additional
characteristic ''charge'' is assigned to the entries of our  sequences
and (\ref{1.10}) is generalized as follows: the ''monochromatic''
segments are broken into subsegments of entries of the same charge so
that entries of larger charge are always on the right-hand side of
entries  of smaller charge (of the same color) and only the entries of
the same charge and color are ordered (nonincreasing from right to
left).

\section {Homogeneous vertex operator construction of basic (level 1 standard)
modules}
\hspace{3ex}Since most of our considerations refer to the vertex
operator algebra approach to the basic $\ga$-modules (the standard
$\ga$-modules of level 1), we shall briefly sketch it in this section.
We work in the setting of [FLM] and [DL], to which we refer for more
details; see also [B], [FK], [K], [S].

   Consider the tensor product vector  spaces $V_{Q} := M(1) \otimes
\C[Q] \;, V_{P} := M(1) \otimes \C[P]$. We shall be using
independent commuting formal variables $z, z_{0}, z_{1}, z_{2},
\ldots \,.$ For any vector space $V$ we denote by $V[[z]]$ the  space
of all (possibly infinite) formal series of nonnegative integral
powers of $z$ with coefficients in $V.$ Similarly, we denote by
$V\{z\}$ the space of all (possibly infinite in both directions)
formal series of rational powers of $z$ with coefficients in $V.$
Recall that $V_{Q}$ has a natural structure of simple vertex operator
algebra (VOA) and that $V_{P}$ is a module for this VOA through a
linear map, which we define on all of $V_{P},$ rather than just
$V_{Q}$ (cf. [DL]):

$$\begin{array}[t]{ccc} V_{P} &\rightarrow &(\mbox{\rm End} V_{P}) \{z \}
\\ v &\mapsto & Y(v,z) \end{array}$$
given by
\begin{equation}
Y(1 \otimes e^{\lambda},z) := \mbox{\rm exp} \left \{ \sum_{n \geq 1}
h_{\lambda} (-n)
\frac{z^{n}}{n} \right \} \, \mbox{\rm exp} \left \{ \sum_{n \geq 1}
h_{\lambda} (n)
\frac{z^{-n}}{-n} \right \} \otimes e^{\lambda} \;
z^{h_{\lambda}}\;\varepsilon_{\lambda}
\label{2.1}\end {equation}
for $\lambda\,\in\, P$ and by

\begin{eqnarray}
&Y(\prod_{i=1}^{l} h_{i}(-n_{i}) \otimes e^{\lambda},\,z ) :=
\nonumber\\ &=\;
 :\prod_{i=1}^{l} \left [ \frac{1}{(n_{i}-1)!} \left (\frac{d}{dz}
\right )^{n_{i}-1} \; h_{i}(z) \right] Y(1 \otimes
e^{\lambda},z):, \label{2.2}\end{eqnarray} for a generic homogeneous
vector $(h_{i}\,\in\,\h,\;n_{i}\,\in\,\Z_{+},\;\lambda\,\in\,P),$
where the following notations are used :$$h(m) := h \otimes t^{m}
\;\;\; \mbox{\rm for all}\; h \in \h ,\, m \in \Z,$$ $$z^{h_{\lambda}}
e^{\mu} := z^{\langle \lambda,\mu\rangle }e^{\mu}
\;\;\;\mbox{for all}\;\;
\lambda, \mu \in P,$$$$\varepsilon_{\lambda} e^{\mu} :=
\varepsilon(\lambda ,\mu) e^{\mu}\;\;\;\mbox{for all}\;\;
\lambda, \mu \in P,$$
$$h(z) := \sum_{m \in \Z} (h(m)\otimes 1) z^{-m-1}\;\;\;\mbox{for
all}\;\;h \in \h,$$ and $: \;\cdot\; :$ is a normal ordering
procedure, which signifies that the enclosed expression is to be
reordered if necessary so that all the operators $h(m)\; (h \in \h,\;
m < 0)$ are to be placed to the left of all the operators $h(m)\;(h
\in \h,\; m \geq 0).$ Abusing notation, we shall often write
$e^{\lambda}$ instead of $1 \otimes e^{\lambda}$ and $h(m)$ instead of
$h(m) \otimes 1$.

   Recall the following classical
interpretation of $V_{P}$ as a $\ga$-module ([FK], [S]): Let  $x_{\alpha}
\otimes t^{m} ,\; \alpha \in \Delta ,\; m \in \Z$ act on
$V_{P}$ as $x_{\alpha}(m)$, where  $x_{\alpha}(m)$ is defined as a
coefficient of the vertex operator $Y(e^{\alpha},z)$ :
\begin{equation}
\sum_{m\in\Z} x_{\alpha}(m)z^{-m-1} := Y(e^{\alpha},z);
\label{2.3}\end{equation}
let $h\otimes t^{m},\; h\in \h, \;m\in \Z$ act on $V_{P}$ as $h(m)$
(recall that this is our abbreviated notation for the operator $h(m)
\otimes 1)$ and finally, let the central element $c$ act as the
identity operator. Then this action endows $V_{Q}$ and
$V_{Q}e^{\Lambda_{j}}\; $ for $\; j=1,\ldots,n$ with the structure of
level 1 standard $\ga$-modules with highest weight vectors
$v(\hat{\Lambda}_{0}) := 1 \otimes 1$ and $v(\hat{\Lambda}_{j}) := 1
\otimes e^{\Lambda_{j}},\; j =1,\ldots,n,$ respectively.  In other
words $ V_{Q} \cong \;L(\hat{\Lambda}_{0}),\; V_{Q}e^{\Lambda_{j}}
\cong L(\hat{\Lambda}_{j} )\;$ for $\; j=1,\ldots,n$ and therefore
$V_{P} \cong \oplus _{j=0}^{n} L(\hat{\Lambda}_{j}).$ Note that from
the very definitions (\ref{2.1}) and (\ref{2.3}), one has
\begin{equation} x_{\alpha}(m) \,e^{\lambda} \;= \; e^{\lambda}\,
x_{\alpha}(m\;+ \; \langle \lambda,\alpha\rangle ), \; \;\lambda \in
P, \alpha \in  \Delta . \label{2.101} \end{equation}

   Adopting the standard notation
\begin{equation}
E^{\pm}(h,z) := \mbox{exp}\left( \sum_{m \geq 1} h(\pm m) \frac{z^{\mp
m}}{\pm m} \right) \label{2.4}\end{equation}
for $h \in \h,$ one can rewrite (\ref{2.1}) as
\begin{equation}
Y(e^{\lambda},z) = E^{-}(-h_{\lambda},z)E^{+}(-h_{\lambda},z)\otimes
e^{\lambda} z^{h_{\lambda}} \varepsilon_{\lambda}.\label{2.5}
\end{equation}

   Recall that the commutation relations among these vertex operators
as well as some nice properties of their products (see  the next section)
follow from the ``commutation relation'' of $E^{+}$ and $E^{-}$ (cf.
Chapter 4 of [FLM])
\begin{equation}
E^{+}(-h_{\lambda},z_{2}) E^{-}(-h_{\mu},z_{1}) = \left(
1-\frac{z_{1}}{z_{2}}\right)^{\langle \lambda , \mu \rangle
}E^{-}(-h_{\mu},z_{1}) E^{+}(-h_{\lambda},z_{2}) , \label{2.6}
\end{equation}
where $\lambda , \mu \in P$ and the binomial expression  is  to be
expanded in nonnegative powers of $\frac{z_{1}}{z_{2}}.$ As a
corollary,
\begin{equation}
E^{+}(-h_{\lambda},z_{2})  E^{-}(-h_{\mu},z_{1}) \otimes (e^{\lambda}
z_{2}^{h_{\lambda}} \varepsilon_{\lambda})(e^{\mu} z_{1}^{h_{\mu}}
\varepsilon_{\mu}) = \label{2.7} \end{equation}
$$= \mbox{const} (z_{2} - z_{1})^{\langle \lambda , \mu
\rangle}E^{-}(-h_{\mu},z_{1})E^{+}(-h_{\lambda},z_{2})\otimes
e^{\lambda + \mu}
z_{2}^{h_{\lambda}}
z_{1}^{h_{\mu}}\varepsilon_{\lambda}\varepsilon_{\mu},$$where
$\mbox{const} \in \C^{\times}.$

   There is a Jacobi identity for the operators $Y(v ,
z),\;v\,\in\,V_{P}$ (see Chapter 5 of [DL]), and we shall be
particularly interested in the following case (see formula (12.5) of [DL]):
For $ \lambda \in Q, \mu \in P; u^{*}, v^{*} \in M(1); \;u := u^{*}
\otimes e^{\lambda}, v := v^{*} \otimes e^{\mu},$ one has
\begin{eqnarray}
&z_{0}^{-1} \delta \left (\frac{z_{1} - z_{2}}{z_{0}} \right
)Y(u,z_{1})Y(v,z_{2}) \;- (-1)^{\langle\lambda, \mu\rangle}c(\lambda
,\mu) z_{0}^{-1} \delta \left (\frac{z_{2} - z_{1}}{-z_{0}} \right
)Y(v,z_{2})Y(u,z_{1}) = \nonumber\\& =\; z_{2}^{-1} \delta \left (
\frac{z_{1} - z_{0}}{z_{2}} \right ) Y(Y(u,z_{0})v, z_{2}),
\label{2.8}\end{eqnarray} where $\delta(z) := \sum _{m \in \Z} z^{m}$
is the usual formal delta function and the binomial expressions are to
be expanded in nonnegative integral powers of the second variable.

   Note that for $\mu \in Q,$ we have $(-1)^{\langle\lambda,
\mu\rangle}c(\lambda
,\mu) = 1,$ giving the ordinary Jacobi identity for the vertex
operator algebra $V_{Q}$ and for its action on the irreducible modules
$V_{Q}e^{\Lambda_{j}},\;j =1,\ldots,n.$ In order to get rid of the
numerical factor $(-1)^{\langle\lambda, \mu\rangle}c(\lambda ,\mu)$
even if $\mu \not\in Q,$ we first note that the fundamental weights
$\Lambda_{1}, \ldots ,\Lambda_{n},$ which are all minuscule,
constitute a set of representatives for the nontrivial cosets of $Q$
in $P.$ As in [DL], formula (12.3), for $\mu \in \Lambda_{j} + Q,\;j=
1,\ldots,n,$ we replace $Y(v,z)$ by ${\cal Y}(v, z) := Y(v, z) e^{i\pi
h_{\Lambda _{j}}} c(\cdot , \Lambda_{j})$ (where the operators
$e^{i\pi h_{\Lambda _{j}}}$ and $c(\cdot , \Lambda_{j})$ are defined
in the obvious ways). For $\mu \in Q$ we simply set ${\cal Y}(v, z) :=
Y(v, z).$ This gives us a linear map $v \mapsto {\cal Y}(v, z)$ {}from
$V_{P}$ to $(\mbox{End}V_{P})\{z\},$ and by Proposition 12.2 of [DL],
we have the ordinary Jacobi identity
\begin{eqnarray}
&z_{0}^{-1} \delta \left (\frac{z_{1} - z_{2}}{z_{0}} \right
)Y(u,z_{1}){\cal Y}(v,z_{2}) \;- z_{0}^{-1} \delta \left (\frac{z_{2}
- z_{1}}{-z_{0}} \right ){\cal Y}(v,z_{2})Y(u,z_{1}) = \nonumber\\& =
z_{2}^{-1} \delta \left ( \frac{z_{1} - z_{0}}{z_{2}} \right ) {\cal
Y}(Y(u,z_{0}) v , z_{2})\label{2.9}\end{eqnarray}(see formula (12.8)
of [DL]). This identity, together with certain other natural
conditions (again see Proposition 12.2 of [DL]), guarantees that
${\cal Y}(\cdot , z)$ defines an intertwining operator in the sense of
[FHL]. More precisely, let us write $\Lambda_{0} := 0$ for
convenience. Then for $\mu \in \Lambda_{j} + Q,\;j= 0, 1,
\ldots,n,$$\;\,{\cal Y}(v , z)$ (with $v$ as above) defines an
intertwining operator of type $\left[\begin{array}{c} l \\j i
\end{array} \right],\; l\, \equiv\,( i+j)\;
\mbox{\rm mod} (n+1),$ since the correspondence $j \mapsto
\Lambda_{j}$ defines an isomorphism from the cyclic group $\Z /(n +
1)\Z$ to $P/Q$ (we are indexing the irreducible $V_{Q}$-modules
$V_{Q}e^{\Lambda_{j}}$ by the integers $j\;\mbox{mod}\;(n+1));$
cf.Chapter 12 of [DL]. In particular, one has a map $${\cal
Y}(e^{\Lambda_{j}},z) : L(\hat{\Lambda}_{i}) \rightarrow
L(\hat{\Lambda}_{l}) \{z \}$$ for $l\,\equiv \, (i + j)\; \mbox{\rm
mod} \;(n+1).$

   Now in (\ref{2.8}) make the specializations $u := e^{\alpha},\;
\alpha \in \Delta_{+} $ and $ v := e^{\Lambda_{j}} $ for any $j \in
\{1,\ldots,n\}$. Notice that from the very definition (\ref{2.1}), one
has $Y(e^{\alpha},z_{0}) e^{\Lambda_{j}} \in V_{P}[[z_{0}]]$ and
therefore, taking $\mbox{\rm Res}_{z_{0}}$ of (\ref{2.8}), we simply
get
\begin{equation}
[Y(e^{\alpha},z_{1}) \;,\;{\cal Y}(e^{\Lambda_{j}},z_{2})] = 0
\;\;\;\;\mbox{for}\;\;
\alpha \in \Delta_{+}, \; j = 1,\ldots,n. \label{2.10}\end{equation}
This seemingly innocent commutativity, which asserts that the
coefficients of the series ${\cal Y}(e^{\Lambda_{j}},z)$ are
$\nb_{+}$-module maps, has deep implications for the representation
theory of $\ga :$ One is tempted to interpret it as a device to
``lift'' relations (or constraints, as physicists would say) between
operators from $U (\na_{+})$ acting on $ L(\hat{\Lambda}_{i}) $ to
analogous relations among these operators, but acting on $
L(\hat{\Lambda}_{l})$. In particular, when questions of basis are
concerned, it can be very advantageous to treat all the simple modules
at a given level simultaneously. We shall later demonstrate fruitful
applications of this strategy (Theorem \ref{The 3.2}).

   We close this Section with the  definition of a projection needed only
for part II [GeII]: Since the grading $\C[Q] = \coprod_{\beta \in Q} \C
e^{\beta}$ of the group lattice $\C[Q]$ induces a
grading of the whole basic module $L(\hat{\Lambda}_{j}) = M(1) \otimes
\C[Q]e^{\Lambda_{j}}, \; 0\leq j \leq n,$ we can  define
\be \pi_{U(\ha^{-})\cdot v(\hat{\Lambda}_{j}) }: L(\hat{\Lambda}_{j} )
\rightarrow
U(\ha^{-})\cdot v(\hat{\Lambda}_{j}) = M(1)\otimes e^{\Lambda_{j}}
\cong M(1) \label{2.12}\ee to be the corresponding projection on the
homogeneous subspace $U(\ha^{-})\cdot v(\hat{\Lambda}_{j}).$ Using the
grading $U(\ha^{-}) = \coprod_{m\in \N}U^{m}(\ha^{-})$ by symmetric
powers $U^{m}(\ha^{-}) = S^{m}(\ha^{-}),$ we can go one step further
and define for every $m \in \N$ the coresponding projection
\be \pi_{U^{m}(\ha^{-})\cdot v(\hat{\Lambda}_{j}) }: L(\hat{\Lambda}_{j} )
\rightarrow
U^{m}(\ha^{-})\cdot v(\hat{\Lambda}_{j}).\label{2.14}\ee

\section{Quasi-particles}
\hspace{3ex}We begin with the choice of a special   subspace of
$U(\nb_{+})$ whose appropriate completions will contain all the
basis-generating operators considered below. Set $U$ to be the
subspace which is the ordered product of universal enveloping algebras
\be U := U(\nb_{\alpha_{n}})  U(\nb_{\alpha_{n-1}}) \cdots
U(\nb_{\alpha_{1}}).\label{5.1}\ee
This is a
product of subalgebras of $U(\ga),$ and by the Poincar\'{e}-Birkhoff-Witt
theorem, it is linearly isomorphic to the tensor product
$U(\nb_{\alpha_{n}})\otimes
U(\nb_{\alpha_{n-1}}) \otimes \cdots \otimes
U(\nb_{\alpha_{1}}).$

   Fix a level $k \in \Z_{+}$ and let
$\hat{\Lambda}$ be a dominant integral highest weight of this level,
$\hat{\Lambda} = k\hat{\Lambda}_{0} + \Lambda,\; \Lambda \in P.$  Denote
the action of $x_{\alpha} \otimes t^{m} \in \ga,$ $\alpha \in \Delta, \; m \in
\Z,$ on the standard module $L(\hat{\Lambda})$ by $x_{\alpha}(m)$ (in the
previous section we have constructed explicitly $x_{\alpha}(m)$ for
basic modules). Denote the corresponding generating function (vertex
operator) by $X_{\alpha}(z),$ i.e.,
\be X_{\alpha}(z):= \sum_{m \in \Z} x_{\alpha}(m) z^{-m -1}, \;
\alpha \in \Delta, \label{5.301}\ee
where $z$ is a formal variable. For example, for the explicit
realization of basic
modules given in  the previous section, one obviously has $X_{\alpha}(z) =
Y(e^{\alpha}, z)$ (cf. (\ref{2.3})).  More generally,  a standard
module  at any level $k \in \Z_{+}$ can be constructed explicitly as a
subspace of  the tensor product of $k$ such basic modules: In this
case we  simply set
$X_{\alpha}(z) := \Delta^{k-1}
(Y(e^{\alpha}, z)),$  where $\Delta^{k-1}$ is the $(k-1)$-iterate of
the  standard coproduct $\Delta,$ and then our module is generated by
the tensor product of the  $k$ level one highest weight vectors (no
confusion can arise from the fact that $\Delta$ denotes also the set
of $\g$-roots).

   Let us quickly show that the principal subspace $W(\hat{\Lambda})$
of the standard module $L(\hat{\Lambda})$ is indeed generated by
operators in $U$ acting on the highest weight vector
$v(\hat{\Lambda})$ (this is true for any highest weight $\ga$-module).

\begin{Lemma}One has
$$W(\hat{\Lambda}) : = U(\nb_{+})\cdot v(\hat{\Lambda}) = U \cdot
v(\hat{\Lambda}).$$\label{Lem 5.1}\end{Lemma}

\noindent{\em Proof} Observe that $W(\hat{\Lambda})$ is spanned by
$$ \{b\cdot v(\hat{\Lambda})| b \in U(\nb_{\beta_{r}}) \cdots
U(\nb_{\beta_{1}});\; \beta_{1},
\ldots,\beta_{r} \,\in\,\Pi\},$$ since every $x_{\beta}, \,\beta\,\in
\,\Delta_{+},$ can be expressed as a bracket of $x_{\alpha_{i}},\;
1\leq i \leq n.$ We want to show that the spanning property will still
hold even if we order this product of universal enveloping algebras.
In other words, we have to find a way to change the order in the
product $x_{\alpha_{i}}(l)x_{\alpha_{i+1}}(m),\;1\leq i \leq n-1,$
when acting on a given vector $v \in W(\hat{\Lambda}),$ possibly at
the expense of changing the indices. But this is easy, because for
every $m\,\in\,\Z,$ there exists $N\,\gg\,0$ such that
$x_{\alpha_{i+1}}(m+N)\cdot v= 0$ and therefore
$$x_{\alpha_{i}}(l)x_{\alpha_{i+1}}(m)\cdot v = \pm
x_{\alpha_{i}+\alpha_{i+1}}(l+m)\cdot v +
x_{\alpha_{i+1}}(m)x_{\alpha_{i}}(l)\cdot v =$$ $$= -
x_{\alpha_{i+1}}(m+N)x_{\alpha_{i}}(l-N)\cdot v +
x_{\alpha_{i+1}}(m)x_{\alpha_{i}}(l)\cdot v.$$\hfill Q.E.D.\vspace{5mm}

    Following physicists' terminology (cf. e.g. [DKKMM1]), we shall
say that an operator $x_{\alpha_{i}}(m)\in \nb_{\alpha_{i}}$
represents a {\em quasi-particle of color $i$ and charge $1$} (the
eigenvalue $-m$ of the scaling operator $D$ under the adjoint action
is called the {\em energy} of our quasi-particle). Moreover, a
monomial from $U(\nb_{\alpha_{n}}) U(\nb_{\alpha_{n-1}})
\cdots U(\nb_{\alpha_{1}})$ is {\em of color-type $(r_{n};r_{n-1};...;r_{1})$}
if it carries charge $r_{n}$ with color $n,$ charge $r_{n-1}$ with
color $n-1$ and so on (we are using the grading of each
$U(\nb_{\alpha_{i}}) = S(\nb_{\alpha_{i}})$ by symmetric powers and
taking the tensor product grading). We thus obtain a ``color-type''
gradation of the whole vector space $U := U(\nb_{\alpha_{n}})
U(\nb_{\alpha_{n-1}})
\cdots U(\nb_{\alpha_{1}}):$
\be U =  \coprod_{r_{n},\ldots ,r_{1} \geq 0}\;
U_{(r_{n};\ldots ;r_{1})}. \label{5.2}\ee Note that for every
(dominant integral) highest weight $\hat{\Lambda},$ the principal
subspace $W(\hat{\Lambda})$ also has a color-type gradation
(compatible with the $\h$-gradation of $L(\hat{\Lambda})):$
\be W(\hat{\Lambda}) = \coprod_{r_{n},\ldots ,r_{1} \geq 0}\;
W(\hat{\Lambda})_{(r_{n};\ldots ;r_{1})},\label{5.302}\ee
where
\be W(\hat{\Lambda})_{(r_{n};\ldots ;r_{1})} := W_{\Lambda +
\sum_{i=1}^{n}r_{i}\alpha_{i}} (\hat{\Lambda})\label{5.303}\ee
is the weight subspace of  weight $\Lambda +
\sum_{i=1}^{n}r_{i}\alpha_{i} \in P.$

   When  a given  monomial is not monochromatic, it is
somewhat  convenient to have the color-type
implicitly encoded in its notation. We shall do this by adding a
second subscript $i$ to
all the entries associated with a quasi-particle of color $i$ and
having
the other subscript  enumerating quasi-particles of a given color only
(as opposed to using  a single subscript as in (\ref{1.10}) for example).

   Unless stated otherwise, we shall assume that a product of
(commuting) quasi-particles of the same color $i$ (and charge $1$) has
its quasi-particle indices nonincreasing from right to left. With this
convention in mind, note that the set of quasi-particle monomials
{}from $U$ of a given color-type $(r_{n};\ldots ;r_{1})$ is linearly
ordered by ''$<$'' if the definition (\ref{1.11}) is applied to the
respective index sequences. Moreover, for two such quasi-particle
monomials $b$ and $b',$ we write $b \prec b'$ if $b < b'$ and in
addition, $$(m_{n}; \ldots ; m_{1}) \prec (m'_{n}; \ldots ; m'_{1})$$
(cf.  (\ref{1.12})), where $m_{i}$ (resp., $m'_{i}),$ $1\leq i \leq
n,$ is the sum of indices of all the quasi-particles of color $i$ in
$b$ (resp., $b').$

   The partial ordering ''$\prec$'' will be pivotal in our spanning
arguments because it has the nice property that $\prec$-intervals are
{\em finite} (which is not true for its linear extension ''$<$''; it
was an idea of J. Lepowsky to employ the multidimensional ordering
''$\prec$'', fundamental in [LW2], in the current setting). The
lexicographic ordering ''$<$'' has this property only if the index-sum
for every single color is held fixed (because quasi-particles of
different colors do not commute).

   One can naturally  generalize the above concept  of quasi-particle
of charge $1$ and define quasi-particles of arbitrary charge $r
\in \Z_{+}$  as follows:

\begin{Definition} For given $i,\; 1\leq i \leq n,$ and $r \in
\Z_{+},$ $m \in \Z,$  define
\be x_{r\alpha_{i}}(m) := \sum_{\begin{array}{c}
{\scriptstyle m_{r},\ldots ,m_{1} \in \Z}\\{\scriptstyle m_{r} +
\cdots +
m_{1} = m }\end{array}} x_{\alpha_{i}}(m_{r})\cdots
x_{\alpha_{i}}(m_{1}) = \label{5.3}\ee $$= \mbox{Res}_{z}\{z^{m + r
-1}\underbrace{X_{\alpha_{i}}( z)\cdots X_{\alpha_{i}}( z)}_{r\;
factors}\}$$ (the indices $m_{r},\ldots ,m_{1}$ in the above multisum
are not ordered!). We call $x_{r\alpha_{i}}(m)$ a {\bf quasi-particle
of color $i$ and charge $r$} (the eigenvalue $-m$ of the scaling
operator $D$ under the adjoint action is as usual the {\em energy} of
our quasi-particle). Abusing language, we shall say that
$x_{r\alpha_{i}}(m)$ is from $U(\nb_{\alpha_{i}})$ because our
quasi-particles will always act on highest weight modules, in which
case, the sum above is finite (note that this sum is infinite in
general, i.e., the quasi-particles lie in an appropriate completion of
$U(\nb_{\alpha_{i}}).$\label{Def 5.1}\end{Definition}\vspace{5mm}

   Note that a quasi-particle of charge $r$ can be thought of as a
cluster of $r$ quasi-particles of charge one  confined  in such a way
that only the total
index-sum (i.e., the energy  of the cluster with
a minus sign) is ``measurable'', while  the individual quasi-particle
indices  run  through $\Z.$  Nevertheless, just like the
quasi-particles of charge $1,$ the  quasi-particles of charge $r$  are
coefficients  of   certain vertex operators: If we set
\be X_{r\alpha_{i}}(z) := \underbrace{X_{\alpha_{i}}( z)\cdots
X_{\alpha_{i}}( z)}_{r\; factors}= \sum_{m \in \Z} x_{r\alpha_{i}}(m)
z^{-m - r},\label{5.101}\ee (cf. (\ref{5.301})), one can show that $
X_{r\alpha_{i}}(z) $ is the vertex operator corresponding to the
vector $x_{\alpha_{i}}(-1)^{r}\cdot v(k\hat{\Lambda}_{0}),$ where
$v(k\hat{\Lambda}_{0})$ is the vacuum highest weight vector at the
chosen level $k \in \Z_{+}.$ In other words,
\be  X_{r\alpha_{i}}(z) = Y(x_{\alpha_{i}}(-1)^{r}\cdot
v(k\hat{\Lambda}_{0}), z), \label{5.102}\ee
(cf. Proposition 13.16 of [DL]).

   Unless stated otherwise, we shall assume that a product of
(commuting) quasi-particles of the same color and charge has its
indices nonincreasing from right to left. Moreover, a product of
(commuting) quasi-particles of the same color but different charges
will have its charges nonincreasing from right to left (not
surprisingly, quasi-particles of the same color but different charges
will behave like different objects; cf. Section 5).

   Pick a Young diagram (partition)
\be r^{(1)} \geq r^{(2)} \geq \cdots \geq r^{(K)} >  0,\; \sum_{t
=1}^{K}r^{(t)} = r,\; K \in
\Z_{+},\label{5.4}\ee
pictured as follows:\\
\setlength{\unitlength}{2pt}
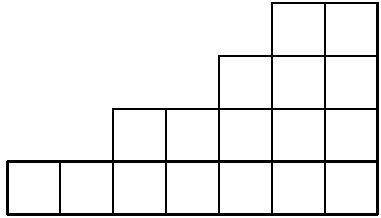
\begin{picture}(100,40)(-30,0)
\put(30,0){\line(1,0){70}}
\put(30,10){\line(1,0){70}}
\put(50,20){\line(1,0){50}}
\put(70,30){\line(1,0){30}} \put(80,40){\line(1,0){20}}
\put(30,0){\line(0,1){10}}\put(40,0){\line(0,1){10}}
\put(50,0){\line(0,1){20}}\put(60,0){\line(0,1){20}}
\put(70,0){\line(0,1){30}}\put(80,0){\line(0,1){40}} \put(90,0){\line(0,1){40}}
\put(100,0){\line(0,1){40}}

\end{picture}
\setlength{\unitlength}{1pt}\vspace{3mm}

\noindent(the bottom row has $r^{(1)}$ squares, the second  row has $r^{(2)}$
squares, $\ldots,$ the top $K^{th}$ row has $r^{(K)}$ squares). Let
the dual Young diagram (in reversed order) be
\be 0< n_{r^{(1)}} \leq \cdots \leq n_{2} \leq n_{1} =  K,\;
\sum_{p=1}^{r^{(1)}} n_{p} = r,\label{5.5}\ee
(i.e., the rightmost column has $n_{1}$ boxes, the second-from-right
column has $n_{2}$ boxes, $\ldots,$ the leftmost column has
$n_{r^{(1)}}$ boxes).

   Fix a color $i, \; 1\leq i \leq n.$ We say that a monochromatic
quasi-particle monomial is of {\em charge-type $(n_{r^{(1)}}, \ldots
,n_{1})$} and of {\em dual-charge-type $(r^{(1)}, \ldots, r^{(K)})$}
if it is built (in the obvious sense) out of $r^{(1)} - r^{(2)}$
quasi-particles of charge $1$, $r^{(2)} - r^{(3)}$ quasi-particles of
charge $2,$ $\ldots,$ $r^{(K)}$ quasi-particles of charge $K.$ In
other words, each quasi-particles of charge $r$ is represented by a
column of height $r$ in the respective Young diagram.

   The same terminology of course applies to the corresponding
generating functions
(this is closer to  the setting  in which Feigin and Stoyanovsky
talk about  clusters; cf. [FS], Theorem 2.7.1): We say that
\be X_{n_{r^{(1)}}\alpha_{i}}(z_{r^{(1)}})\cdots
X_{n_{1}\alpha_{i}}(z_{1}) \label{5.6}\ee is of {\em charge-type
$(n_{r^{(1)}}, \ldots ,n_{1})$} and of {\em dual-charge-type
$(r^{(1)}, \ldots, r^{(K)})$}.

   In complete analogy with the charge one  picture,  we can extend
our dictionary to multi-colored  quasi-particle monomials from $U$
\be b := b_{n}\cdots b_{2}b_{1},\label{5.10}\ee
where $b_{i}, \; 1\leq i \leq n,$ is a monochromatic quasi-particle
monomial of color $i$ (simply order the colors and add a subscript $i$
to the entries corresponding to the color $i).$ It is clear what it
means for the quasi-particle monomial (\ref{5.10}) to be of {\em
color-charge-type}
\be (n_{r_{n}^{(1)},n},\ldots ,n_{1,n};\ldots ; n_{r_{1}^{(1)},1},
\ldots , n_{1,1}),\label{5.11}\ee
where
$$0 < n_{r^{(1)},i} \leq \cdots \leq n_{2,i} \leq n_{1,i} \leq
K,\;\sum_{p =1}^{r_{i}^{(1)}} n_{p,i} = r_{i},\; 1\leq i \leq n,$$
of {\em color-dual-charge-type}
\be (r_{n}^{(1)}, \ldots , r_{n}^{(K)}; \ldots ; r_{1}^{(1)},
\ldots , r_{1}^{(K)}),\label{5.12}\ee
where
$$ r_{i}^{(1)} \geq r_{i}^{(2)} \geq \cdots \geq r_{i}^{(K)} \geq 0,\; \sum_{t
=1}^{K}r_{i}^{(t)} = r_{i},\; K \in
\Z_{+},\;1\leq i \leq n$$
(we shall sometimes be writing only the leftmost and the rightmost
entries inside the parentheses) and of {\em color-type} $(r_{n};
\ldots; r_{1}).$  It shall also say that the corresponding generating function
\be X_{n_{r_{n}^{(1)},n}\alpha_{n}}(z_{n_{r_{n}^{(1)},n}}) \cdots
X_{n_{1,1}\alpha_{1}}(z_{1,1})\label{5.13}\ee is of the above
color-charge-type and color-dual-charge-type.

   Finally, we can naturally extend both the linear ordering ''$<$''
and the partial ordering ''$\prec$'' to the set of multi-colored
higher-charge quasi-particle monomials of given color-type
$(r_{n};\ldots ;r_{1}).$ The lexicographic ordering ''$<$'' is defined
as follows: First apply definition (\ref{1.11}) to the {\em
color-charge-types} of the two monomials $b$ and $b'$ (rather than to
their index sequences!); if the color-charge-types are the same, apply
(\ref{1.11}) to the index sequences of the two monomials. In complete
analogy with the charge-one situation, the partial ordering
''$\prec$'' is defined by ''restricting'' the lexicographic ordering
''$<$'' as follows: We write $b \prec b'$ if $b < b'$ and in addition,
$$(m_{n}; \ldots ; m_{1}) \prec (m'_{n}; \ldots ; m'_{1})$$ (cf.
(\ref{1.12})), where $m_{i}$ (resp., $m'_{i}),$ $1\leq i \leq n,$ is
the sum of indices of all the quasi-particles of color $i$ in $b$
(resp., $b').$

   One should be aware that throughout the rest of the paper, the
lexicographic ordering in the set of color-charge-types (for fixed
color-type) will be truncated from above in a very special way.
Namely, formula (\ref{5.102}) and the null vector identity
\be x_{\alpha_{i}}(-1)^{k+1}\cdot
v(k\hat{\Lambda}_{0}) = 0,\label{5.15}\ee (where
$v(k\hat{\Lambda}_{0})$ is the vacuum highest weight vector at a given
level $k \in \Z_{+}),$ imply that
\be X_{(k+1)\alpha_{i}}(z) = \underbrace{X_{\alpha_{i}}(z)\cdots
X_{\alpha_{i}}( z)}_{k+1\; factors} = Y(x_{\alpha_{i}}(-1)^{k+1}\cdot
v(k\hat{\Lambda}_{0}),z) = 0,\label{5.201}\ee i.e., all the
quasi-particles of charge greater than $k$ are zero when acting on
level $k$ standard modules (this was the main idea on which [LP2] was
based).  ``Annihilating'' relations like this are not enough for
constructing a quasi-particle basis at levels $k > 1.$ We shall have
to employ in addition the following (obvious from (\ref{5.101}) and
(\ref{5.102})) relations which express quasi-particle monomials of a
given (color)-charge-type through quasi-particle monomials of the same
color-type but of greater in the ordering ''$<$''
(color)-charge-types: for $0 <s \leq s' \leq k,$ one has
\be X_{s\alpha_{i}}(z)X_{s'\alpha_{i}}(z) =
X_{(s-1)\alpha_{i}}(z)X_{(s'+1)\alpha_{i}}(z) =\label{5.16}\ee
$$= \cdots \cdots = X_{\alpha_{i}}(z)X_{(s'+s-1)\alpha_{i}}(z) =
X_{(s'+s)\alpha_{i}}(z).$$
These are $s$ independent relations for  monochromatic
quasi-particle
monomials of charge-type $(s,s'),\; 0 < s \leq s' \leq k,$ which
express them through  quasi-particle
monomials  of  greater charge-types. One
can actually rewrite
these  relations among vertex operators as (equally obvious)
equivalent relations among the corresponding vectors, cf. (\ref{5.102}):
\be x_{s\alpha_{i}}(-s) x_{s'\alpha_{i}}(-s')\cdot
v(k\hat{\Lambda}_{0}) = x_{(s-1)\alpha_{i}}(-(s-1))
x_{(s'+1)\alpha_{i}}(-(s'+1))\cdot
v(k\hat{\Lambda}_{0})  =\label{5.17}\ee
$$= \cdots\cdots = x_{\alpha_{i}}(-1)x_{(s'+s-1)\alpha_{i}}(-(s'+s-1)\cdot
v(k\hat{\Lambda}_{0}) =  x_{(s'+s)\alpha_{i}}(-(s'+s))\cdot
v(k\hat{\Lambda}_{0}) = $$
$$= \underbrace{x_{\alpha_{i}}(-1) x_{\alpha_{i}}(-1)\cdots
x_{\alpha_{i}}(-1)}_{s'+s\, factors}\cdot
v(k\hat{\Lambda}_{0}).$$

   Another series of fundamental relations, independent from the ones
above,  is the
following: for $0 < s < s' \leq k,$ one has
\be \frac{1}{s}\left(\frac{d}{dz}X_{s\alpha_{i}}(z)\right)
X_{s'\alpha_{i}}(z) = \frac{1}{s'}
X_{s\alpha_{i}}(z)\left(\frac{d}{dz}
X_{s'\alpha_{i}}(z)\right),\label{5.18}\ee which follows from the
corresponding vector relation
\be \frac{1}{s}x_{s\alpha_{i}}(-s-1)x_{s'\alpha_{i}}(-s')\cdot
v(k\hat{\Lambda}_{0}) =
\frac{1}{s'}x_{s\alpha_{i}}(-s)x_{s'\alpha_{i}}(-s'-1)\cdot
v(k\hat{\Lambda}_{0}) = \label{5.19}\ee
$$= \underbrace{x_{\alpha_{i}}(-2) x_{\alpha_{i}}(-1) \cdot
x_{\alpha_{i}}(-1)}_{s'+s\, factors}\cdot
v(k\hat{\Lambda}_{0})  .$$
Note that the relation (\ref{5.18}) is trivial for $s = s'.$ Combining
(\ref{5.18}) with (\ref{5.16}), we can replace (\ref{5.18}) by another
set of relations, independent from (\ref{5.16}): for
$0 < s \leq s'
\leq k,$ one has
\be \frac{d}{dz}\left(X_{s\alpha_{i}}(z)X_{s'\alpha_{i}}(z)\right)  =
\frac{s+s'}{s}\left(\frac{d}{dz}X_{s\alpha_{i}}(z)\right)
X_{s'\alpha_{i}}(z) = \label{5.20}\ee $$=
\frac{s+s'}{s-1}\left(\frac{d}{dz}X_{(s-1)\alpha_{i}}(z)\right)
X_{(s'+1)\alpha_{i}}(z) = \cdots \cdots\cdots \cdots\cdots =$$ $$=
\frac{s+s'}{1}\left(\frac{d}{dz}X_{\alpha_{i}}(z)\right)
X_{(s'+s-1)\alpha_{i}}(z)$$ (for $0 < s < s' \leq k$ these are $s$ new
independent relations).  The vector counterparts of these relations
are
\be  x_{s\alpha_{i}}(-s-1)x_{s'\alpha_{i}}(-s')\cdot
v(k\hat{\Lambda}_{0}) +
x_{s\alpha_{i}}(-s)x_{s'\alpha_{i}}(-s'-1)\cdot
v(k\hat{\Lambda}_{0})  = \label{5.21}\ee
$$= \frac{s+s'}{s}x_{s\alpha_{i}}(-s-1)x_{s'\alpha_{i}}(-s')\cdot
v(k\hat{\Lambda}_{0}) = $$
$$ =
\frac{s+s'}{s-1}x_{(s-1)\alpha_{i}}(-(s-1)-1)x_{(s'+1)\alpha_{i}}(-(s'+1))\cdot
v(k\hat{\Lambda}_{0}) = \cdots \cdots \cdots =$$ $$ = \frac{s + s'}{1}
x_{\alpha_{i}}(-1)x_{(s'+s-1)\alpha_{i}}(-(s'+s-1))\cdot
v(k\hat{\Lambda}_{0}).$$

\section{Quasi-particle basis for the principal subspaces of basic
modules}

\hspace{3ex}Throughout this section we shall simply talk about
quasi-particles without specifying their charge since they all will
have charge $1.$

   Consider all the $n+1$ basic modules $L(\hat{\Lambda}_{j}),$ $
0\leq j \leq n,$ as constructed in Section 2, i.e., set
\be X_{\beta}(z) := Y(e^{\beta}, z),\; \beta \in \Delta,
\label{3.301}\ee
(cf. (\ref{2.1})-(\ref{2.3})).

   For every $j,
\; 0 \leq j \leq n, $ we propose a basis for the corresponding
principal subspace $W(\hat{\Lambda}_{j}) \subset $
$L(\hat{\Lambda}_{j}). $ Here is the set of quasi-particle monomials
which generate our basis (when acting on the highest weight vector
$v(\hat{\Lambda}_{j})):$

\begin{Definition} {\rm Fix $j, \; 0 \leq j \leq n.$ Set

\be \frak{B}_{W(\hat{\Lambda}_{j})} := \bigsqcup_{r_{n},\ldots , r_{1} \geq
0}\label{3.1}\ee $$\left\{ x_{\alpha_{n}}(m_{r_{n},n})\cdots
x_{\alpha_{n}}(m_{1,n})\cdots \cdots x_{\alpha_{1}}(m_{r_{1},1})
\cdots x_{\alpha_{1}}(m_{1,1})
\left | \begin{array} {l}\\ \\ \\ \end{array}\right.\right.$$
$$\left|\begin{array}{l} m_{p,i} \in \Z,\; 1\leq i \leq n, \; 1 \leq
p \leq r_{i};\\m_{p,i} \leq  r_{i-1} -
\delta_{i,j} - 2(p-1) - 1;\\m_{p+1,i} \leq m_{p,i} -2 \end{array}\right\},$$
where $r_{0} := 0$ (when $r_{i} = 0,\; 1\leq i \leq n,$ no
quasi-particles of color $i$
are present). }\label{Def 3.1}\end{Definition}\vspace{5mm}
The first  of the two
nontrivial conditions condition is a
truncation condition which in particular incorporates the interaction
between quasi-particles of color $i$ and quasi-particles of color $i-1$
(quasi-particles of colors
$i$ and $i- p,\; p >1,$ do not interact with each other). The second
condition is nothing else but the
``difference two at distance one'' condition for the quasi-particles
of the same color $i$ (cf. Preliminaries).\vspace{5mm}

\noindent{\em Example 4.1} Consider $\g = sl(3)$ (i.e., $n=2)$  and the vacuum
principal subspace $W(\hat{\Lambda}_{0})$ (i.e., $j = 0).$ We shall
denote for brevity the monomial $x_{\alpha_{2}}(s) \cdots
x_{\alpha_{1}}(t)$ by $(s_{\alpha_{2}} \ldots t_{\alpha_{1}}).$ For
the first few energy levels (the eigenvalues of the scaling operator
$D$ under the adjoint action), we list in Table 1 of the Appendix the
elements of $\frak{B}_{W(\hat{\Lambda}_{0})}$ of color-types $(1;2)$
and $(2;2).$ \vspace{5mm}

   Note that due to the second condition in the above definition, we
can weaken the  first  condition and still get the same set:
\be \frak{B}_{W(\hat{\Lambda}_{j})} = \bigsqcup_{r_{n},\ldots , r_{1} \geq 0}
\left\{ x_{\alpha_{n}}(m_{r_{n},n})\cdots
x_{\alpha_{n}}(m_{1,n})\cdots \cdots x_{\alpha_{1}}(m_{r_{1},1}) \cdots
x_{\alpha_{1}}(m_{1,1})
\left | \begin{array} {l}\\ \\ \\ \end{array}\right.\right.\label{3.101}\ee
$$\left|\begin{array}{l} m_{p,i} \in \Z,\; 1\leq i \leq n, \; 1 \leq
p \leq r_{i};\\m_{p,i} \leq  r_{i-1} -
\delta_{i,j} - 1;\\m_{p+1,i} \leq m_{p,i} -2 \end{array}\right\}.$$

   The first nontrivial condition  in (\ref{3.101}) is encoded in the
assertion  of the following lemma.

\begin{Lemma}Fix $j, \;0\leq j \leq n.$ One has
\be \left[ \prod_{i =2}^{n} \prod_{p=1}^{r_{i}}
\prod_{q =1}^{r_{i-1}} \left( 1  - \frac{z_{q,i-1}}{z_{p,i}}\right) \right]
X_{\alpha_{n}}(z_{r_{n},n})
\cdots  X_{\alpha_{1}}(z_{1,1}) \cdot v(\hat{\Lambda}_{j}) \in
\label{3.3} \ee
$$\in \left[\prod_{i=1}^{n} \prod_{p=1}^{r_{i}}
z_{p,i}^{\delta_{i,j} - r_{i-1}}
\right] W(\hat{\Lambda}_{j}) [[z_{r_{n},n}, \ldots ,
z_{1,1}]]$$
where $r_{0} := 0.$

\label{Lem 3.1}\end{Lemma}

\noindent {\em Proof} Follows from a more general claim:
\be \left[ \prod_{\begin{array}{c} {\scriptstyle
i,l=1}\\{\scriptstyle i \geq l}\end{array}}^{n} \prod_{p=1}^{r_{i}}
\prod_{\begin{array}{c}{\scriptstyle q=1}\\{\scriptstyle p > q\; for\;
i =l}\end{array}}^{r_{l}} (z_{p,i} - z_{q,l})^{-\langle \alpha_{i},
\alpha_{l}\rangle} \right] X_{\alpha_{n}}(z_{r_{n},n}) \cdots
X_{\alpha_{1}}(z_{1,1}) \cdot v(\hat{\Lambda}_{j}) \in \label{3.4}
\ee
$$\in \left[\prod_{i=1}^{n} \prod_{p=1}^{r_{i}} z_{p,i}^{\delta_{i,j}}
\right] W(\hat{\Lambda}_{j}) [[z_{r_{n},n}, \ldots ,
z_{1,1}]],$$ where the binomial expressions are to be expanded as
usual in nonnegative integral powers of the second variable. But this
claim is an immediate implication of the fact that $\ha^{+} \cdot
v(\hat{\Lambda}_{j}) = 0$ and the commutation relation (\ref{2.7}),
applied to $\lambda = \alpha_{l}, \;\mu = \alpha_{i},\;z_{2} =
z_{q,l},\; z_{1} = z_{p,i}$ (see also (\ref{2.4}) and (\ref{2.5})).
\hfill Q.E.D.  \vspace{5mm}

   One should not fail to observe that on the left-hand  side of
(\ref{3.4}) we have simply removed ``by hand'' all the ``universal''
poles and zeroes in the generating  function
$$X_{\alpha_{n}}(z_{r_{n},n}) \cdots
X_{\alpha_{1}}(z_{1,1}) \cdot v(\hat{\Lambda}_{j}).$$
Namely, these are  the (order one) poles on the
hyperplanes
\be z_{p,i} = z_{q,i-1},\;2\leq i \leq n,\; 1\leq p \leq r_{i},\; 1\leq
q \leq r_{i-1}\label{3.5}\ee
and the (order two) zeroes on the hyperplanes
\be z_{p,i} = z_{q,i},\;1\leq i \leq n,\;1 \leq q < p \leq
r_{i}.\label{3.6}\ee
These singularities ought to be ``blamed'' for the interaction between
the quasi-particles. It is clear how  to identify the restricted dual to
$W(\hat{\Lambda}_{j})$ space with an appropriate space of symmetric
(with respect to each group of variables $z_{r_{i},i},\ldots ,
z_{1,i})$
polynomials and thus make  a  connection with
the Feigin-Stoyanovsky construction [FS].

   We are now all set for a quick demonstration of the spanning
property of the declared basis.

\begin{Theorem} For  a given $j, \; 0\leq j \leq n,$ the set $\{b\cdot
v(\hat{\Lambda}_{j})\;|\;b \in
\frak{B}_{W(\hat{\Lambda}_{j})}\} $ spans the principal subspace
$W(\hat{\Lambda}_{j})$ of $L(\hat{\Lambda}_{j}).$ \label{The
3.1}\end{Theorem}

\noindent{\em Proof} In view of Lemma \ref{Lem 5.1}, it suffices to
show that every vector $b\cdot v(\hat{\Lambda}_{j}),$ $b$ a monomial
in $U,$ is a linear combination of the proposed vectors.

   If a monomial $b$ of color-type $(r_{n};\ldots ;r_{1})$ violates
the condition
\be m_{p,i} \leq r_{i-1} -
\delta_{i,j} -1\label{3.103}\ee
for some $i, \; 1\leq i \leq n,$ and $p,\;1 \leq p \leq r_{i},$ then
Lemma \ref{Lem 3.1} implies that the vector $b\cdot
v(\hat{\Lambda}_{j})$ is a linear combination of vectors of the form
$b'\cdot v(\hat{\Lambda}_{j}),$ $b'$ a monomial from $U,$ with $b'$
and $b$ having the same color-type and {\em total} index-sum and $b'
\succ b$ (but there is at least one color $i$ for which the
corresponding index-sums in $b'$ and $b$ are different; cf.
Section 3 where the ordering ''$\prec$'' for quasi-particle monomials was
introduced). There are only finitely many such monomials $b'$ which do
not annihilate $v(\hat{\Lambda}_{j}).$

   On the other hand, (\ref{5.201}) with $k=1$ furnishes a new
independent constraint for every pair of quasi-particles of the same
color. It implies that if a monomial $b$ from $U$ violates the
condition
\be m_{p+1,i} \leq m_{p,i} -2 \label{3.104}\ee
(the ``difference two at distance one `` condition for a given color
$i,\; 1 \leq i \leq n),$ we can express
 the vector
$b\cdot v(\hat{\Lambda}_{j})$ as a linear combination of
vectors of the form  $b' \cdot v(\hat{\Lambda}_{j}),$ $b'$ a
monomial from $U,$ with $b'$ and $b$ having  the same color-type
and the same index-sum for
every given color $i$ and  in addition,  $b' \succ b$ (note again that
there are only finitely many
such monomials $b'$ which do not annihilate $v(\hat{\Lambda}_{j})).$

   We have shown that if a monomial violates either of the two
nontrivial conditions in definition (\ref{3.101}), it can be raised in
the ordering ''$\prec$'' (which has finite intervals). In other words,
after finitely many steps, we can express every vector $b\cdot
v(\hat{\Lambda}_{j}),$ $b$ a monomial from $U,$ through vectors from
the proposed set $\{b \cdot v(\hat{\Lambda}_{j})\;|b \in
\frak{B}_{W(\hat{\Lambda}_{j})}\}.$   \hfill  Q.E.D.
\vspace{5mm}

\noindent{\em Remark 4.1} We have actually proven that every vector
$b\cdot v(\hat{\Lambda}_{j}),$ $b$ a quasi-particle monomial
{}from $U,\;b
\not\in \frak{B}_{W(\hat{\Lambda}_{j})},$
is a linear combination of vectors of the form $b' \cdot
v(\hat{\Lambda}_{j}),$ $b' \in \frak{B}_{W(\hat{\Lambda}_{j})},$ with
$b'$ and $b$ having the same color-type and total index-sum and
moreover $b' \succ b.$ \vspace{5mm}

   As suggested in Section 2, we are going to prove the independence
of the above spanning set using intertwining operators between
different modules (although there are other possible approaches, for
example, through the dual picture of Feigin and Stoyanovsky [FS]).
\begin{Theorem} For a given $j,\; 0\leq j \leq n,$ the set $\{ b\cdot
v(\hat{\Lambda}_{j})|b\in
\frak{B}_{W(\hat{\Lambda}_{j})} \}$ forms  a basis for  the principal subspace
$W(\hat{\Lambda}_{j})$ of $L(\hat{\Lambda}_{j}).$\label{The 3.2}
\end{Theorem}

\noindent{\em Proof} We shall   prove first
that a monomial relation $b\cdot v(\hat{\Lambda}_{j})=  0,$ $b
\in \frak{B}_{W(\hat{\Lambda}_{j})},$ would imply
$v(\hat{\Lambda}_{j}) =  0 $ and
therefore is impossible. This will only show that each proposed basis
vector  is nonzero but a slight elaboration of this argument will
later eliminate the possibility for {\em any} linear relation among
the vectors and thus prove their independence.

   Choose for
concreteness a monomial of color-type $(r_{n};\ldots ; r_{2};r_{1})$
\be  b  := x_{\alpha_{n}}(m_{r_{n},n})
\cdots x_{\alpha_{2}}(m_{1,2}) x_{\alpha_{1}}(m_{r_{1},1})
\cdots x_{\alpha_{1}}(m_{1,1}) \in  \frak{B}_{W(\hat{\Lambda}_{j})}
\label{3.201}\ee
and assume that a relation $b\cdot v(\hat{\Lambda}_{j})= 0$ holds.
Apply
\be {\rm Res}_{z}\left(z^{-1- \langle \Lambda_{1},
\Lambda_{j}\rangle}{\cal Y}(e^{\Lambda_{1}},z)\right)\label{3.202}\ee
on both sides of the relation and employ (\ref{2.10}) to move this
operator all the way to the right. Then use
\be {\rm Res}_{z}\left(z^{-1- \langle \Lambda_{1},
\Lambda_{j}\rangle}{\cal Y}(e^{\Lambda_{1}},z)\right)\cdot
v(\hat{\Lambda}_{j}) = \mbox{const}\,e^{\Lambda_{1}} \cdot
v(\hat{\Lambda}_{j}),\;\; \mbox{const} \in \C^{\times},\label{3.203}\ee
and (\ref{2.101}) to move back the operator $e^{\Lambda_{1}}$ all the
way to the left  at the expense of increasing by one the indices of
all the quasi-particles of color 1. Drop the invertible
operator $e^{\Lambda_{1}}$ and conclude that $b' \cdot
v(\hat{\Lambda}_{j})=  0,$ where
\be b' =   x_{\alpha_{n}}(m_{r_{n},n})
\cdots x_{\alpha_{2}}(m_{1,2}) x_{\alpha_{1}}(m_{r_{1},1}+1)
\cdots x_{\alpha_{1}}(m_{1,1}+1)  \in  \frak{B}_{W(\hat{\Lambda}_{j})}.
\label{3.204}\ee
Repeat the same trick until the rightmost index reaches its maximal
allowed value (before the corresponding quasi-particle is annihilated
by the highest weight vector), namely,
$m_{1,1} = -1 - \delta_{1,j}.$ Since
\be x_{\alpha_{1}}(-1- \delta_{1,j})\cdot v(\hat{\Lambda}_{j}) =
\mbox{const}\,e^{\alpha_{1}}\cdot
v(\hat{\Lambda}_{j}),\;\; \mbox{const} \in \C^{\times},\label{3.205}\ee
formula (\ref{2.101}) allows to move the operator
$e^{\alpha_{1}}$ all the way to
the left at the expense of increasing by two  the indices of
all the quasi-particles of color 1 and decreasing by one the indices of
all the quasi-particles of color 2. Dropping the invertible operator
$e^{\alpha_{1}},$ conclude that $b'' \cdot
v(\hat{\Lambda}_{j})=  0,$ where
\be  b'' =   x_{\alpha_{n}}(m_{r_{n},n})
\cdots x_{\alpha_{2}}(m_{1,2} - 1) x_{\alpha_{1}}(m_{r_{1},1}+2)
\cdots x_{\alpha_{1}}(m_{2,1}+2) \in \frak{B}_{W(\hat{\Lambda}_{j})}
\label{3.206}\ee
is of color-type $(r_{n};\ldots ;r_{2}; r_{1} - 1)$. Repeat this whole
cycle and decrease the number of quasi-particles until none of them is
left, i.e., the false identity $v(\hat{\Lambda}_{j}) = 0$ is obtained
-- contradiction!

   Now assume that a
general  linear relation  $\sum_{s=0}^{m}
\,\xi_{s}b_{s} \cdot v(\hat{\Lambda}_{j})= 0 $ holds, $\xi_{s} \;\in
\;\C^{\times },\;
b_{s} \in \frak{B}_{W(\hat{\Lambda}_{j})}$ all distinct (we can assume
that the constraint is not a sum of two nontrivial constraints and
this implies that all $b_{s}$ are of the same color-type). Execute the
above reduction procedure for the monomial which is smallest (among
those involved) in the linear lexicographic ordering ''$<$'' (cf.
(\ref{1.11}) and Section 3) and enjoy the observation that -- by the
very definition of ''$<$'' -- all the other monomials get annihilated
at some intermediate stage of the reduction (because they are reduced
to the form $ \cdots x_{\alpha_{i}}(m)$ for some $m > -1 -
\delta_{i,j}, \;1 \leq i \leq n ).$ So, the outcome is again the false
identity $v(\hat{\Lambda}_{j}) = 0 $ -- contradiction! \hfill  Q.E.D.
\vspace{5mm}

   It is truly remarkable that the above  recursive reduction actually
fails if we start
with a monomial $ b$ from $U$ which  is not in
$\frak{B}_{W(\hat{\Lambda}_{j})}$ (in other words we get at some stage of the
induction  to the trivial but true identity $ 0 = 0$). This means
that we could have discovered our basis just from the requirement that
the above {\em reductio ad absurdum}  works!

   Without any further elaborations, we can write down character
formulas for the principal subspaces. The only two observations needed
are the fundamental combinatorial identity (cf. [A])
\be \frac{1}{(q)_{r}}:= \frac{1}{(1-q)(1-q^{2})\cdots (1-q^{r})} =
\label{3.11}\ee
$$= \sum_{m \geq 0}\left\{\begin{array}{c}\mbox{number of partitions
of $m$ with}\\\mbox{at most $r$ parts}\end{array}\right\}\,q^{m},$$
where $r \in \Z_{+},\; (q)_{0} := 1,$ and the simple numerical
identity
\be \sum_{p=1}^{r} (2(p-1) + 1) = 1+3 + 5 + \cdots +(2r -1) =
r^{2}.\label{3.12}\ee
{}From the very Definition \ref{Def 3.1}, we immediately  conclude that for
every $j,\; 0 \leq j \leq n,$ one has
\be \mbox{\rm Tr}\,q^{D} \left| \begin{array}[t]{l} \\
W(\hat{\Lambda}_{j}) \end{array} \right.  =  \label{3.13}\ee
$$= \sum_{r_{1} \geq 0}
\frac{q^{r_{1}^{2} + r_{1}\delta_{1,j}}}{(q)_{r_{1}}}\,\sum_{r_{2} \geq 0}
\frac{q^{r_{2}^{2} + r_{2}(\delta_{2,j}  -
r_{1})}}{(q)_{r_{2}}}\,\cdots\,\sum_{r_{n}  \geq 0}
\frac{q^{r_{n}^{2}  + r_{n}(\delta_{n,j} - r_{n - 1})}}{(q)_{r_{n}}}.$$
Equivalently,
\be \mbox{\rm Tr}\,q^{D} \left| \begin{array}[t]{l} \\
W(\hat{\Lambda}_{j}) \end{array} \right.  = \label{3.14}\ee $$=
\sum_{r_{1},\ldots ,r_{n} \geq 0} \frac{q^{\frac{1}{2}\sum_{l,m =
1}^{n} A_{lm}r_{l}r_{m}}}{\prod_{i=1}^{n}(q)_{r_{i}} }\;q^{r_{j}},$$
where $r_{0}:= 0,$ and $(A_{lm})_{l,m = 1}^{n}$ is the Cartan matrix
of $\g = sl(n + 1, \C).$

   In the case of the vacuum module $(j = 0 ),$
this is the Feigin-Stoyanovsky character formula [FS].

   In terms of a
generating  function which
encodes the quasi-particle structure of the basis,  one has
\be  \mbox{ch}
W(\hat{\Lambda}_{j})= \label{3.15}\ee $$= \sum_{r_{1},\ldots ,r_{n}
\geq 0} \frac{q^{\frac{1}{2}\sum_{l,m = 1}^{n}
A_{lm}r_{l}r_{m}}}{\prod_{i=1}^{n}(q)_{r_{i}}
}\;q^{r_{j}}\prod_{i=1}^{n} y_{i}^{r_{i}}.$$ The coefficient of
$y_{1}^{r_{1}}\cdots y_{n}^{r_{n}}$ on the right-hand side gives the
$q$-character of the weight subspace $W_{\Lambda_{j} +
\sum_{i=1}^{n}r_{i}
\alpha_{i}}(\hat{\Lambda}_{j}).$

\section{Quasi-particle  basis for the principal subspaces at any positive
integral level $k$}

\hspace{3ex}Fix a level $k \in  \Z_{+} $. We shall consider for
simplicity only highest weights of the form
\be \hat{\Lambda} :=
k_{0}\hat{\Lambda}_{0} + k_{j}\hat{\Lambda}_{j} = k \hat{\Lambda}_{0} +
\Lambda\;\;\mbox{where}\;\Lambda := k_{j}\Lambda_{j},\label{4.201}\ee
for some $j,\; 1 \leq j \leq n;$ $k_{0}, k_{j} \in \N $ and $k_{0} +
k_{j} = k.$ (All dominant integral weights are of this form if $n=1,$
i.e., $\g = sl(2,\C)$ or if $k=1.)$ We would like to warn the reader
that this restriction is not as innocuous as it might look: neither
Definition \ref{Def 4.1} of our basis, nor the subsequent statements
(e.g., the crucial for the spanning argument Lemma \ref{Lem 4.1}) are
immediately generalizable for other highest weights.

   It is convenient to define
\be  j_{t} :=  \left\{ \begin{array}{l} \ 0\;\; \mbox{for} \;
0 < t \leq  k_{0} \\  j\;\;  \mbox{for} \;
k_{0}< t \leq k = k_{0}+ k_{j}.\end{array}
\right.\label{4.1}\end{equation}

   Simulating the level one bases built in the previous section, we
shall propose a basis for the principal subspace $W(\hat{\Lambda})$
$\subset$ $L(\hat{\Lambda}),$ which will be generated by
quasi-particles (of charge no greater than $k)$ acting on the highest
weight vector $v(\hat{\Lambda}).$ Not surprisingly, our main technical
tool will be the realization of $W(\hat{\Lambda})$ as a subspace of
the tensor product of $k$ level one modules. More precisely,
\be W(\hat{\Lambda}) =
U(\nb_{+}) \cdot v(\hat{\Lambda})\subset
W(\hat{\Lambda}_{j_{k}})\otimes \cdots \otimes
W(\hat{\Lambda}_{j_{1}})\subset  V_{P}^{ \otimes\;k}\label{4.2}\ee
where
\be
v(\hat{\Lambda}) := v(\hat{\Lambda}_{j_{k}})\otimes \cdots \otimes
v(\hat{\Lambda}_{j_{1}}) = \label{4.3}\ee
$$=\underbrace{v(\hat{\Lambda}_{j})\otimes\cdots \otimes
v(\hat{\Lambda}_{j})}_{k_{j}\;factors} \otimes
\underbrace{v(\hat{\Lambda}_{0})\otimes\cdots \otimes
v(\hat{\Lambda}_{0})}_{k_{0}\;factors}$$
and $U(\nb_{+})$ acts through  $\Delta^{k-1}$ (the $(k-1)$-fold
iterate of the standard
coproduct $\Delta$ in the bialgebra $U(\nb_{+});$ $\Delta^{0} :=
\mbox{id});$  no confusion can arise from the fact that  $\Delta$
denotes also the set of $\g$-roots).  In other words, we set
\be X_{\beta}(z) := \Delta^{k-1}(Y(e^{\beta},z)) = \label{4.101}\ee
$$\underbrace{Y(e^{\beta}, z)\otimes \mbox{\bf 1}\otimes  \cdots
\otimes\mbox{\bf 1}}_{k\,factors} + \underbrace{\mbox{\bf 1}\otimes
Y(e^{\beta}, z) \otimes \cdots
\otimes\mbox{\bf 1}}_{k\,factors} + \cdots + \underbrace{\mbox{\bf
1}\otimes \cdots
\otimes\mbox{\bf 1}\otimes Y(e^{\beta}, z)}_{k\,factors},\; \beta \in
\Delta. $$
Note that by Lemma \ref{Lem 5.1}, one has $W(\hat{\Lambda}) = U \cdot
v(\hat{\Lambda}).$

   We shall freely use all the notations, definitions and formulas
from Section 3.

   Below is the set of quasi-particle monomials from $U$ which
generate our basis (see the subsequent example). We keep the format of
Definition \ref{Def 3.1} in order to help the reader to get through
this confusing abundance of indices (we shall alleviate the pain by
pinpointing the correspondence between the entries in Definition
\ref{Def 3.1} and their generalizations in the current context). Our
set is a disjoint union over color-charge--types (cf. Section 3)
\be (n_{r_{n}^{(1)},n},\ldots ,n_{1,n};\ldots ; n_{r_{1}^{(1)},1},
\ldots , n_{1,1}),\label{4.4}\ee
$$0 < n_{r_{i}^{(1)},i} \leq \cdots \leq n_{2,i} \leq n_{1,i} \leq
k,\;\sum_{p =1}^{r_{i}^{(1)}} n_{p,i} =: r_{i},\; 1\leq i \leq n,$$
or, equivalently, over  the color-dual-charge-types
\be (r_{n}^{(1)}, \ldots , r_{n}^{(k)}; \ldots ; r_{1}^{(1)},
\ldots , r_{1}^{(k)}),\label{4.5}\ee
$$ r_{i}^{(1)} \geq r_{i}^{(2)} \geq \cdots \geq r_{i}^{(k)} \geq 0,\;
\sum_{t =1}^{k}r_{i}^{(t)} = r_{i},\;1\leq i \leq n$$ ($r_{i}^{(1)}$
distinct quasi-particles of color $i$ and charge at most $k$ are
present).

\begin{Definition} Fix a highest weight $\hat{\Lambda}$ as in
(\ref{4.201}). Set
\be \frak{B}_{W(\hat{\Lambda})} :=
\bigsqcup_{\begin{array}{c}{\scriptstyle 0 \leq
n_{r_{n}^{(1)},n} \leq \cdots \leq  n_{1,n}\leq k}\\{\scriptstyle
\cdots\cdots\cdots}\\{\scriptstyle 0 \leq
n_{r_{1}^{(1)},1} \leq \cdots \leq n_{1,1}\leq
k}\end{array}}\left(\mbox{or,
equivalently,}\;\;\bigsqcup_{\begin{array}{c}{\scriptstyle
r_{n}^{(1)}\geq \cdots \geq r_{n}^{(k)}\geq 0}\\{\scriptstyle
\cdots\cdots\cdots}\\{\scriptstyle
r_{1}^{(1)} \geq \cdots \geq  r_{1}^{(k)}\geq
0}\end{array}}\right)
\label{4.6}\ee
$$\left\{x_{n_{r_{n}^{(1)},n}\alpha_{n}}(m_{r_{n}^{(1)},n})
\cdots x_{n_{1,n}\alpha_{n}}(m_{1,n}) \cdots \cdots
x_{n_{r_{1}^{(1)},1}\alpha_{1}}(m_{r_{1}^{(1)},1}) \cdots
x_{n_{1,1}\alpha_{1}}(m_{1,1})\begin{array}{c}\\ \\ \\
\end{array}\right|$$
$$\left|\begin{array}{ll} m_{p,i} \in \Z,\; 1\leq i \leq n, \; 1 \leq
p \leq r_{i}^{(1)};&\\m_{p,i} \leq
\sum_{q=1}^{r_{i-1}^{(1)}} \mbox{min}\,\{n_{p,i}, n_{q,i-1}\}  -
\sum_{t=1}^{n_{p,i}}\delta_{i,j_{t}}
-\sum_{p>p'>0} 2\mbox{min}\, \{n_{p,i}, n_{p',i}\} -
n_{p,i};&\\m_{p+1,i} \leq m_{p,i} -2n_{p,i}\;\;\mbox{for}\;\;
n_{p+1,i} = n_{p,i}& \end{array}\right\}, $$ where $r_{0}^{(1)} := 0$
and $j_{t}$ was defined in (\ref{4.1}). \label{Def
4.1}\end{Definition}\vspace{5mm}

   It is rather obvious  that the role of $r_{i}$ (the total charge
due to color $i)$ in Definition \ref{Def 3.1} is played here by
$$\sum_{q=1}^{r_{i}^{(1)}} n_{q,i} = \sum_{t =1}^{k}r_{i}^{(t)} =:
r_{i}$$ and moreover, a quasi-particle $x_{\alpha_{i}}(m_{p,i})$ of
charge $1$ from Definition \ref{Def 3.1} is simply replaced here by a
quasi-particle $x_{n_{p,i}\alpha_{i}}(m_{p,i})$ of charge $n_{p,i}.$
The ``difference two at distance one'' condition for quasi-particles
of the same color and charge $1$ in Definition \ref{Def 3.1} is
generalized in the current setting to a ``difference $2n_{p,i}$ at
distance one'' condition for quasi-particles of the same color and
charge $n_{p,i}.$ Not surprisingly, the quasi-particles of the same
color, but of {\em different} charge, have to be treated as different
objects since they do not satisfy any reasonable difference conditions
among themselves (although they commute). The first (truncation)
condition in Definition \ref{Def 3.1} has its entries (on the
right-hand side of the inequality) generalized here as follows:
$$\begin{array}{ccc}r_{i-1}= \sum_{q=1}^{r_{i-1}} 1\; &\mapsto &\;
\sum_{q=1}^{r_{i-1}^{(1)}}
\mbox{min}\,\{n_{p,i}, n_{q,i-1}\} \\& & \\ - \delta_{i,j}\;&\mapsto &\;
- \sum_{t=1}^{n_{p,i}}\delta_{i,j_{t}}\\& & \\ - 2(p-1) =
-\sum_{p>p'>0} 2\;&\mapsto &\;
-\sum_{p>p'>0} 2\mbox{min}\, \{n_{p,i}, n_{p',i}\} = -\sum_{p>q>0}
2 n_{p,i} \\& & \\ -1\;&\mapsto &\;-
n_{p,i}. \end{array}$$\vspace{5mm}

\noindent{\em Example 5.1} Consider $\g \;=\; sl(3)$ (i.e.,
$n=2),$ $ k =2$
and the vacuum principal subspace $W(2\hat{\Lambda}_{0}).$ We shall
denote for brevity the quasi-particle monomial $x_{s'\alpha_{2}}(s)
\cdots x_{t'\alpha_{1}}(t)$ by $(s_{s'\alpha_{2}} \ldots
t_{t'\alpha_{1}}).$ For the first few energy levels (the eigenvalues
of the scaling operator $D$ under the adjoint action), we list in
Table 2 of the Appendix the elements of
$\frak{B}_{W(2\hat{\Lambda}_{0})}$ of color-types $(1;2)$ and
$(2;2).$ \vspace{5mm}

   It is illuminating to have the entries in this truncation  condition
 written down in terms of the color-dual-charge-type parameters
$r_{i}^{(t)}.$ Suppose  $n_{p,i} = s,\; 1\leq s \leq k.$ Then
\be \sum_{q=1}^{r_{i-1}^{(1)}}
\mbox{min}\,\{n_{p,i}, n_{q,i-1}\} = \sum_{q=1}^{r_{i-1}^{(1)}}
\mbox{min}\,\{s, n_{q,i-1}\} =
\sum_{t=1}^{s}r_{i-1}^{(t)}.\label{4.7}\ee
Since the number of quasi-particles  of charge $s$ and color $i$ is
$r_{i}^{(s)} - r_{i}^{(s+1)},$ the total ``shift'' due to the
interaction between quasi-particles  of colors $i$ and $i-1$ is
\be \sum_{s=1}^{k}(r_{i}^{(s)} -
r_{i}^{(s+1)})\sum_{t=1}^{s}r_{i-1}^{(t)} = \label{4.8}\ee $$=
r_{i}^{(1)}r_{i-1}^{(1)} + r_{i}^{(2)}r_{i-1}^{(2)} + \cdots +
r_{i}^{(k)}r_{i-1}^{(k)}$$ (this and the subsequent identities will be
needed later for the character formulas which are written in terms of
$r_{i}^{(s)}).$ Similarly, the total ``shift'' due to the delta
functions is
\be -\sum_{t=1}^{k}r_{i}^{(t)} \delta_{i,j_{t}}.\label{4.9}\ee
A longer but straightforward calculation  shows that the total ``shift''
due to the interaction between quasi-particles  of the same color $i$ is
\be - \sum_{p=1}^{r_{i}^{(1)}}\left( \sum_{p>p'>0} 2\mbox{min}\,
\{n_{p,i}, n_{p',i}\} + n_{p,i}\right) =
-\sum_{\begin{array}{c}{\scriptstyle p,p' = 1}\\{\scriptstyle p >
p'}\end{array}}^{r_{i}^{(1)}} 2n_{p,i} - \sum_{t=1}^{k} r_{i}^{(t)} =
\label{4.10} \ee
$$= - \sum_{t=1}^{k} r_{i}^{(t)^{2}}$$
(cf. [FS], Theorem 2.7.1).

   Similarly to the transition between
Definition \ref{Def 3.1} and the equivalent definition (\ref{3.101}), we
can weaken the first  condition in Definition \ref{Def 4.1}, omitting
the terms on the right-hand side which are due to the
interaction between quasi-particles  of the same color and charge (the second
condition implies that the original inequalities automatically hold).
Since for $n_{p,i} = s,$ one has
\be  -\sum_{n_{p',i} > n_{p,i}} 2\mbox{min}\, \{n_{p,i}, n_{p',i}\} =
-\sum_{n_{p',i} > n_{p,i}} 2n_{p,i} = -2s \sum_{t=
s+1}^{k}r_{i}^{(t)},\label{4.11}\ee
the set $\frak{B}_{W(\hat{\Lambda})}$ can be alternatively described
as follows (cf.
also (\ref{4.7})):
\be \frak{B}_{W(\hat{\Lambda})} =
\bigsqcup_{\begin{array}{c}{\scriptstyle 0 \leq
n_{r_{n}^{(1)},n} \leq \cdots \leq n_{1,n}\leq k}\\{\scriptstyle
\cdots\cdots\cdots}\\{\scriptstyle 0 \leq
n_{r_{1}^{(1)},1} \leq \cdots \leq  n_{1,1}\leq
k}\end{array}}\left(\mbox{or,
equivalently,}\;\;\bigsqcup_{\begin{array}{c}{\scriptstyle
r_{n}^{(1)}\geq  \cdots \geq  r_{n}^{(k)}\geq 0}\\{\scriptstyle
\cdots\cdots\cdots}\\{\scriptstyle
r_{1}^{(1)} \geq \cdots \geq  r_{1}^{(k)}\geq
0}\end{array}}\right)
\label{4.12}\ee
$$\left\{x_{n_{r_{n}^{(1)},n}\alpha_{n}} (m_{r_{n}^{(1)},n})
\cdots x_{n_{1,n}\alpha_{n}}(m_{1,n}) \cdots \cdots
x_{n_{r_{1}^{(1)},1}\alpha_{1}}(m_{r_{1}^{(1)},1}) \cdots
x_{n_{1,1}\alpha_{1}}(m_{1,1})\begin{array}{c}\\ \\ \\ \\
\end{array}\right|$$
$$\left|\begin{array}{ll} m_{p,i} \in \Z,\; 1\leq i \leq n, \; 1 \leq
p \leq r_{i}^{(1)};&\\
m_{p,i} \leq \sum_{q=1}^{r_{i-1}^{(1)}} \mbox{min}\,\{n_{p,i}, n_{q,i-1}\}  -
\sum_{t=1}^{n_{p,i}}\delta_{i,j_{t}}
-\sum_{n_{p',i} > n_{p,i}} 2n_{p,i} - n_{p,i} = \\ =
\sum_{t=1}^{s}r_{i-1}^{(t)} -
\sum_{t=1}^{s}\delta_{i,j_{t}}  - 2s \sum_{t=
s+1}^{k}r_{i}^{(t)} - s, \;\;\mbox{where}\; s:= n_{p,i};&\\m_{p+1,i} \leq
m_{p,i}
-2n_{p,i}\;\;\mbox{for}\;\; n_{p+1,i} = n_{p,i}& \end{array}\right\}.
$$\vspace{5mm}
In contrast to the level one basis
(\ref{3.101}), we have here  a new term
\be -\sum_{n_{p',i} > n_{p,i}}
2n_{p,i} = - 2s \sum_{t=
s+1}^{k}r_{i}^{(t)}\label{4.104}\ee
(whith no  analog in (\ref{3.101})),
 incorporating the interaction between a given  quasi-particle of
charge $n_{p,i} = s$ and all
the quasi-particles   of the same color but greater charges.

   The following Lemma is the higher level generalization of Lemma
\ref{Lem 3.1}.

\begin{Lemma}Fix a highest weight $\hat{\Lambda}$ as in (\ref{4.201}). For a
generating function
$$X_{n_{r_{n}^{(1)},n}\alpha_{n}} (z_{r_{n}^{(1)},n})
\cdots  X_{n_{1,1}\alpha_{1}}(z_{1,1})$$
of color-charge-type
$$ (n_{r_{n}^{(1)},n},\ldots ,n_{1,n};\ldots ; n_{r_{1}^{(1)},1},
\ldots , n_{1,1})$$
(cf. (\ref{4.4})) and a corresponding color-dual-charge-type
$$ (r_{n}^{(1)}, \ldots , r_{n}^{(k)}; \ldots ; r_{1}^{(1)},
\ldots , r_{1}^{(k)}),$$
(cf. (\ref{4.5})), one has
\be \left[ \prod_{i =2}^{n} \prod_{p=1}^{r_{i}^{(1)}}
\prod_{q =1}^{r_{i-1}^{(1)}} \left( 1  -
\frac{z_{q,i-1}}{z_{p,i}}\right)^{\mbox{\rm min} \{n_{p,i},n_{q,i-1}\}} \right]
X_{n_{r_{n}^{(1)},n}\alpha_{n}} (z_{r_{n}^{(1)},n})
\cdots  X_{n_{1,1}\alpha_{1}}(z_{1,1}) \cdot v(\hat{\Lambda}) \in
\label{4.13} \ee
$$\in \left[\prod_{i=1}^{n} \prod_{p=1}^{r_{i}^{(1)}}
z_{p,i}^{\sum_{t=1}^{n_{p,i}}\delta_{i,j_{t}} -
\sum_{q=1}^{r_{i-1}^{(1)}}
\mbox{\rm min}\,\{n_{p,i}, n_{q,i-1}\}}\right] W(\hat{\Lambda})
[[z_{r_{n}^{(1)},n}, \ldots , z_{1,1}]],$$
where $r_{0}^{(1)} := 0.$
\label{Lem 4.1}\end{Lemma}

\noindent{\em Proof}  Follows immediately from Lemma \ref{Lem 3.1},
(\ref{5.201}) with $k=1$ and the explicit form (\ref{4.101}) of the
iterated coproduct $\Delta^{k-1}.$ Note that we are able to encode the
highest weight $\hat{\Lambda}$ in the simple term
$\sum_{t=1}^{n_{p,i}}\delta_{i,j_{t}}$ only because of the special
choice (\ref{4.201}). \hfill  Q.E.D. \vspace{5mm}

   We are now ready to state the higher level generalization of the
spanning Theorem \ref{The 3.1}. We continue our strategy of imitating
the level one picture. The only essentially new element in the
proof below (as compared to the proof of Theorem \ref{The 3.1}) is the
separate treating of the new term $-\sum_{n_{p',i} > n_{p,i}}
2n_{p,i},$ incorporating the interaction between
quasi-particles  of the same color but  different
charges (cf. (\ref{4.12}) and (\ref{4.104})).

\begin{Theorem} For  a given highest weight $\hat{\Lambda}$ as in
(\ref{4.201}), the
set $\{b\cdot v(\hat{\Lambda})\;|\;b \in
\frak{B}_{W(\hat{\Lambda})}\} $ spans the principal subspace
$W(\hat{\Lambda})$ of
$L(\hat{\Lambda}).$ \label{The 4.1}\end{Theorem}

\noindent{\em Proof} Since Lemma \ref{Lem 5.1} holds for any level, it
suffices to
show that every vector $b\cdot v(\hat{\Lambda}),$ $b$ a quasi-particle
monomial {}from $U,$ is a linear combination of the proposed vectors.

   Suppose a quasi-particle monomial $b$ of color-charge-type
(\ref{4.4}) (and a corresponding dual-color-charge-type (\ref{4.5}))
violates the condition
\be m_{p,i} \leq
\sum_{q=1}^{r_{i-1}^{(1)}} \mbox{min}\,\{n_{p,i}, n_{q,i-1}\}  -
\sum_{t=1}^{n_{p,i}}\delta_{i,j_{t}} - n_{p,i} = \label{4.15}\ee
$$=  \sum_{t=1}^{s}r_{i-1}^{(t)}  -
\sum_{t=1}^{s}\delta_{i,j_{t}}  - s,$$
where $s:= n_{p,i},$ and $1\leq i \leq n,$ $1 \leq p \leq r_{i}^{(1)}$
(this is the first nontrivial condition in (\ref{4.12}) with the term
$-\sum_{n_{p',i} > n_{p,i}} 2n_{p,i}$ dropped). Then Lemma \ref{Lem
4.1} implies that the vector $b\cdot v(\hat{\Lambda})$ is a linear
combination of vectors of the form $b'\cdot v(\hat{\Lambda}),$ $b'$ a
quasi-particle monomial from $U,$ $b' \succ b,$ with $b'$ and $b$
having the same color-charge-type and {\em total} index-sum (but there
is at least one color $i$ for which the corresponding index-sums in
$b'$ and $b$ are different). There are only finitely many such
quasi-particle monomials $b'$ which do not annihilate
$v(\hat{\Lambda}).$

   Now the constraints (\ref{5.16}) and (\ref{5.20}) come to action.
They furnish $2s$ new independent (nontrivial) relations for
monochromatic quasi-particle monomials of (color)-charge-type
$(s,s'),\, 0<s < s'\leq k.$ Some of these relations involve
quasi-particle monomials of the same (color)-type, but of greater
(color)-charge-types. Adding these new relations, we can strengthen
the inequality from the last paragraph and claim that if a
quasi-particle monomial $b$ of color-charge-type (\ref{4.4}) violates
the stronger condition
\be m_{p,i} \leq
\sum_{q=1}^{r_{i-1}^{(1)}} \mbox{min}\,\{n_{p,i}, n_{q,i-1}\}  -
\sum_{t=1}^{n_{p,i}}\delta_{i,j_{t}} -\sum_{n_{p',i} > n_{p,i}}
2n_{p,i} - n_{p,i} = \label{4.16}\ee
$$=  \sum_{t=1}^{s}r_{i-1}^{(t)}  -
\sum_{t=1}^{s}\delta_{i,j_{t}}- 2s \sum_{t=
s+1}^{k}r_{i}^{(t)} - s,$$ where $s:= n_{p,i},$ and $1\leq i \leq n,$
$1 \leq p \leq r_{i}^{(1)}$ (this is exactly the first nontrivial
condition in (\ref{4.12})), then the vector $b\cdot v(\hat{\Lambda})$
is still a linear combination of vectors $b' \cdot v(\hat{\Lambda}),$
$ b'$ a quasi-particle monomial {}from $U,\;b'\succ b,$ with $b'$ and
$b$ having the same color-type and {\em total} index-sum (but the
color-charge-type is typically different; no need to say that there
are only finitely many such quasi-particle monomials $b'$ which do not
annihilate $v(\hat{\Lambda})).$ In order to see this, one has to be
more patient than usual and induct on the number $\sum_{t=
s+1}^{k}r_{i}^{(t)}$ of quasi-particles of color $i$ and charge
greater than $s = n_{p,i}$ (this is exactly the number of summands in
the new term (\ref{4.104}) which distinguishes (\ref{4.16}) from its
predecessor (\ref{4.15})). Namely, using the $2n_{p,i}$ new
constraints for our quasi-particle $x_{n_{p,i}\alpha_{i}}(m_{p,i})$
and its closest right neighbor of the same color $i$ and greater
charge, one obtains vectors of the desired form and (finitely many)
other vectors of the form $b'' \cdot v(\hat{\Lambda}),$ where $b''$ is
of the same color-charge-type as $b$ but  its quasi-particle
$x_{n_{p,i}\alpha_{i}}$ has index $\geq m_{p,i} + 2n_{p,i}.$ It
remains to use the inductive assumption for each of the vectors $b''
\cdot v(\hat{\Lambda})$ and observe that they are all expressible
through vectors $b' \cdot v(\hat{\Lambda})$ of the desired form, i.e.,
such that not only $b' \succ b'',$ but also $b'
\succ b!$

   Finally, the constraints (\ref{5.16}) furnish $s$ new independent
(nontrivial) relations for monochromatic quasi-particle monomials of
(color)-charge-type $(s,s),\, 0< s \leq k.$ This means that if a
quasi-particle monomial $b$ of color-charge-type (\ref{4.4}) violates
the ``difference $2n_{p,i}$ at distance one'' condition
\be m_{p+1,i} \leq m_{p,i}
-2n_{p,i}\;\;\mbox{for}\;\; n_{p+1,i} = n_{p,i}=: s,\label{4.17}\ee
(this is the second nontrivial condition in (\ref{4.12})), then the
vector $b\cdot v(\hat{\Lambda})$ is a linear combination of vectors of
the form $b' \cdot v(\hat{\Lambda}),$ $b'$ a quasi-particle monomial
{}from $U,\;b'\succ b,$ with $b'$ and $b$ having the same color-type
and index-sum for any given color $i$ (but the color-charge-type might
be different; there are again only finitely many such quasi-particle
monomials $b'$ which do not annihilate $v(\hat{\Lambda})).$

      Together with the conclusion of  the previous paragraph and formula
(\ref{4.12}), this
guarantees that -- after finitely many steps --   we
can express every  vector $b\cdot v(\hat{\Lambda}),$ $b$ a quasi-particle
monomial from  $U,$
through vectors from the proposed set $\{b \cdot v(\hat{\Lambda})\;|b \in
\frak{B}_{W(\hat{\Lambda})}\}.$   \hfill Q.E.D.
\vspace{5mm}

\noindent{\em Remark 5.1} The proof implies that Remark 4.1 is
true for  level $k$ highest weights:  Every vector
$b\cdot v(\hat{\Lambda}),$  $b$ a quasi-particle  monomial
{}from $U,\;b
\not\in \frak{B}_{W(\hat{\Lambda})},$  is a linear combination of
vectors of the form $b' \cdot v(\hat{\Lambda}),$ $b' \in
\frak{B}_{W(\hat{\Lambda})},\; b' \succ b$ with  $b'$ and $b$ having    the
same color-type and total index-sum. \vspace{5mm}

   We proceed with the definition  of a projection  needed
for generalizing  the independence Theorem \ref {The 3.2} to
level $k.$

   Consider the direct sum decomposition
\be W(\hat{\Lambda}_{j_{k}})\otimes \cdots \otimes W(\hat{\Lambda}_{j_{1}})  =
\coprod_{\begin{array}{c}{\scriptstyle
r_{n}^{(k)}, \ldots , r_{1}^{(k)} \geq 0}\\\cdots \cdots
\cdots\\{\scriptstyle r_{n}^{(1)}, \ldots , r_{1}^{(1)} \geq
0}\end{array}} W(\hat{\Lambda}_{j_{k}})_{(r_{n}^{(k)};\ldots
;r_{1}^{(k)})} \otimes \cdots \otimes
W(\hat{\Lambda}_{j_{1}})_{(r_{n}^{(1)};\ldots ;r_{1}^{(1)})},
\label{4.18}\ee (cf. (\ref{5.302}) and ((\ref{4.1})). For a chosen
color-charge-type (\ref{4.4}) and corresponding color-dual-charge-type
$$ (r_{n}^{(1)}, \ldots , r_{n}^{(k)}; \ldots ; r_{1}^{(1)},
\ldots , r_{1}^{(k)}) $$
(cf. (\ref{4.5})), set
\be \pi_{(r_{n}^{(1)};\ldots ; r_{1}^{(k)})}\;:\;
W(\hat{\Lambda}_{j_{k}})\otimes \cdots \otimes
W(\hat{\Lambda}_{j_{1}}) \rightarrow
W(\hat{\Lambda}_{j_{k}})_{(r_{n}^{(k)};\ldots ;r_{1}^{(k)})} \otimes
\cdots \otimes W(\hat{\Lambda}_{j_{1}})_{(r_{n}^{(1)};\ldots
;r_{1}^{(1)})}\label{4.19}\ee to be the projection given by the above
decomposition (we shall denote by the same letter the obvious
generalization of this projection to the space of formal series with
coefficients in $W(\hat{\Lambda}_{j_{k}})\otimes \cdots \otimes
W(\hat{\Lambda}_{j_{1}})).$ For a generating function of the chosen
color-charge-type, one can now conclude from the definition
(\ref{4.101}) and the constraint (\ref{5.201}) (with $k=1)$ that
\be \pi_{(r_{n}^{(1)};\ldots ;
r_{1}^{(k)})}\cdot X_{n_{r_{n}^{(1)},n}\alpha_{n}}(z_{r_{n}^{(1)},n}) \cdots
X_{n_{1,1}\alpha_{1}}(z_{1,1})\cdot v(\hat{\Lambda}) = \label{4.22}\ee
$$= \mbox{const}\,Y(e^{\alpha_{n}},z_{r_{n}^{(k)},n})\cdots
Y(e^{\alpha_{n}},z_{1,n})\cdots\cdots Y(e^{\alpha_{1}},z_{r_{1}^{(k)},1})\cdots
Y(e^{\alpha_{1}},z_{1,1})\cdot v(\hat{\Lambda}_{j_{k}}) \otimes $$
$$\otimes \cdots\cdots \cdots \cdots \otimes $$
$$\otimes  Y(e^{\alpha_{n}},z_{r_{n}^{(1)},n})\cdots
Y(e^{\alpha_{n}},z_{1,n})\cdots\cdots Y(e^{\alpha_{1}},z_{r_{1}^{(1)},1})\cdots
Y(e^{\alpha_{1}},z_{1,1})\cdot v(\hat{\Lambda}_{j_{1}}) ,$$
where $\mbox{const} \in
\C^{\times}$(a tensor product of $k$ factors).  In order to see this,
fix a color $i,\; 1\leq i \leq n,$ and first ''accommodate'' the
$n_{1,i}$ vertex operators $Y(e^{\alpha_{i}},z_{1,i})$ whose product
generates the $i$-colored quasi-particles of charge $n_{1,i}$ (the
greatest charge for color $i$ in our monomial) -- the projection $\pi$
forces them to spread only along the $n_{1,i}$ rightmost tensor slots
and in addition, (\ref{5.201}) (with $k=1)$ ensures that at most one
(and hence, exactly one) vertex operator is apllied on each of these
tensor slots. Proceed with the remaining vertex operators of color $i$
in the very same fashion. Therefore, for a given quasi-particle
monomial $b$ of the above type, the projection
$\pi_{(r_{n}^{(1)};\ldots ; r_{1}^{(k)})}\cdot b \cdot
v(\hat{\Lambda})$ is a sum of tensor products of $k$
charge-one-quasi-particle monomials (acting on $v(\hat{\Lambda})),$
such that every quasi-particle of charge $s$ from $b$ has exactly one
representative (level one quasi-particle of charge 1) on each of the
$s$ rightmost tensor slots and only there. To put it in different
words, when a color $i$ is fixed, one might associate the $s^{th}$
tensor slot (counted from right to left) of the above tensor product
with the $s^{th}$ row (counted from the bottom to the top) of the
Young diagram in Section 3, filling each box with a level one
quasi-particle of charge one.

   We can now take on the independence of the proposed basis vectors,
generalizing  the level one independence proof (Theorem \ref{The
3.2}). The projection $\pi$ introduced above is  needed  to ensure
that if a level one intertwining  operators ''shuttles'' along  the
$s^{th}$ tensor slot (counted from right to left) as in the proof of
Theorem \ref{The 3.2}, it can  shift the
indices of the quasi-particles of charge $s$ (and the same color) without
affecting quasi-particles  of  smaller charges. The reader is advised to  look
back at the proof of  Theorem \ref{The 3.2}, since we shall only
sketch  here the modifications needed to carry out the
argument in the present  setting.

\begin{Theorem} For  a given highest weight $\hat{\Lambda}$ as in
(\ref{4.201}), the
set $\{b\cdot v(\hat{\Lambda})| b \in
\frak{B}_{W(\hat{\Lambda})}\} $ is indeed a basis for the principal
subspace $W(\hat{\Lambda})$ of
$L(\hat{\Lambda}).$  \label{The 4.2}\end{Theorem}

\noindent{\em Proof} Pick a quasi-particle  monomial $b$ from
$\frak{B}_{W(\hat{\Lambda})}, $
\be b := x_{n_{r_{n}^{(1)},n}\alpha_{n}}(m_{r_{n}^{(1)},n})
\cdots x_{n_{1,n}\alpha_{n}}(m_{1,n}) \cdots \cdots
x_{n_{r_{1}^{(1)},1}\alpha_{1}}(m_{r_{1}^{(1)},1}) \cdots
x_{n_{1,1}\alpha_{1}}(m_{1,1})\label{4.23}\ee of color-charge-type
$$(n_{r_{n}^{(1)},n},\ldots ,n_{1,n};\ldots ; n_{r_{1}^{(1)},1},
\ldots , n_{1,1})$$
(cf. (\ref{4.4})) and corresponding color-dual-charge-type
$$(r_{n}^{(1)}, \ldots , r_{n}^{(k)}; \ldots ; r_{1}^{(1)},
\ldots , r_{1}^{(k)}) = (r_{n}^{(1)}, \ldots , r_{n}^{(k)}; \ldots ;
r_{1}^{(1)},
\ldots , r_{1}^{(s)},0,\ldots ,0),\;\; s:=n_{1,1}, $$
(cf. (\ref{4.5})). Assume that $b\cdot v(\hat{\Lambda}) =
0$ and hence, $\pi_{(r_{n}^{(1)};\ldots ;
r_{1}^{(k)})} \cdot b\cdot v(\hat{\Lambda}) = 0.$  Shuttling back and forth
along the left-hand side of the last identity
with the operator
\be {\bf 1}\otimes \cdots \otimes {\bf 1} \otimes {\rm
Res}_{z}\left(z^{-1- \langle \Lambda_{1},\Lambda_{j_{s}}\rangle}{\cal
Y}(e^{\Lambda_{1}},z)\right)\otimes \underbrace{{\bf 1}\otimes \cdots
\otimes {\bf 1}}_{s-1\,factors},\label{4.24}\ee
increase the indices of all the quasi-particles of charge $s =
n_{1,1}$ and color 1 until the index of the rightmost such
quasi-particle reaches its maximal allowed value (before this
quasi-particle gets annihilated by the highest weight vector), namely,
$m_{1,1} = -s -
\sum_{t=1}^{s}\delta_{1,j_{t}}$ (cf. the proof of Theorem \ref{The
3.2}). Note that none of the quasi-particles in
$\pi_{(r_{n}^{(1)};\ldots ; r_{1}^{(k)})} \cdot b\cdot
v(\hat{\Lambda})$ of smaller charge or different color is affected by
these operations and all the newly obtained vectors are still
projections of quasi-particle monomials from
$\frak{B}_{W(\hat{\Lambda})}$ acting on $v(\hat{\Lambda}).$ In other
words, we end up with the vector
\be \pi_{(r_{n}^{(1)};\ldots ;
r_{1}^{(k)})}\cdot b'\,x_{s\alpha_{1}}(-s -
\sum_{t=1}^{s}\delta_{1,j_{t}})\cdot v(\hat{\Lambda}) =\label{4.25}\ee
$$= \mbox{const}\,\pi_{(r_{n}^{(1)};\ldots ;
r_{1}^{(k)})}\cdot b'\left( {\bf
1}\otimes \cdots \otimes {\bf 1} \otimes
\underbrace{e^{\alpha_{1}}\otimes \cdots \otimes
e^{\alpha_{1}}}_{s\,factors}\right) \cdot v(\hat{\Lambda})$$
for some $\mbox{const} \in \C^{\times}$ and  a quasi-particle
monomial $b' \in  \frak{B}_{W(\hat{\Lambda})}$
of color-charge-type
$$(n_{r_{n}^{(1)},n},\ldots ,n_{1,n};\ldots ; n_{r_{1}^{(1)},1},
\ldots , n_{2,1})$$
and corresponding color-dual-charge-type $$(r_{n}^{(1)}, \ldots ,
r_{n}^{(k)}; \ldots ; r_{1}^{(1)}-1,
\ldots , r_{1}^{(s)}-1,0 \ldots , 0)$$
(i.e., the rightmost quasi-particle of color $1$ and charge $n_{1,1}$
is not present in $b'').$
 Now formula (\ref{2.101}) allows to move the invertible operator
$${\bf
1}\otimes \cdots \otimes {\bf 1} \otimes
\underbrace{e^{\alpha_{1}}\otimes \cdots \otimes
e^{\alpha_{1}}}_{s\,factors}$$ all the way to the left and drop it.
The result is the relation $\pi_{(r_{n}^{(1)},\ldots ,
r_{1}^{(s)}-1,0,\ldots ,0)} \cdot b'' \cdot v(\hat{\Lambda}) = 0$ for
a quasi-particle monomial $b'' \in \frak{B}_{W(\hat{\Lambda})} $ with
one quasi-particle less (and of the same color-charge-type as $b').$
Keep decreasing the number of quasi-particles in the very same fashion
until none of them is left, i.e., the false identity $v(\hat{\Lambda})
= 0$ is obtained -- contradiction!

   Finally, if a general linear relation holds, execute the above
reduction for the quasi-particle monomial, minimal in the linear
lexicographic ordering ''$<$'' (we can assume without loss of
generality that all the quasi-particle monomials involved are of the
same color-type and have the same total index-sum).  During the
process, the quasi-particle monomials with greater (in ''$<$'')
color-charge-types are eliminated due to the projection $\pi$ and the
truncation (\ref{5.201}) (with $k=1$), while the quasi-particle
monomials of the same color-charge-type, but with greater (in ''$<$'')
index sequence, are annihilated by the vacuum vector just like in the
level one picture. The result is again the false identity
$v(\hat{\Lambda}) = 0$ -- contradiction!
\hfill Q.E.D.
\vspace{5mm}

   As expected, the above reasoning works {\em only} for our very
special quasi-particle monomials from $\frak{B}_{W(\hat{\Lambda})}$.
Indeed, it was our ambition to employ such an independence argument
that served as a heuristic for discovering the basis-generating set
$\frak{B}_{W(\hat{\Lambda})}!$

   Let us devote our final effort to writing down a  character
formula for $W(\hat{\Lambda})$ corresponding to the above basis. From the
very Definition \ref{Def 4.1} and  (\ref{3.11}), (\ref{4.8}),
(\ref{4.9}), (\ref{4.10}), one has for a highest weight $\hat{\Lambda}$ as in
(\ref{4.201}) the following character formula:

\be \mbox{\rm Tr}\,q^{D} \left| \begin{array}[t]{l} \\
W(\hat{\Lambda}) \end{array} \right.  = \label{4.26}\ee $$=
\sum_{r_{1}^{(1)}\geq \ldots \geq r_{1}^{(k)} \geq 0}\;
\frac{q^{{r_{1}^{(1)}}^{2}+ \ldots + {r_{1}^{(k)}}^{2} +
\sum_{t=1}^{k} r_{1}^{(t)}\delta_{1,j_{t}} }}{(q)_{r_{1}^{(1)} -
r_{1}^{(2)}} \ldots (q)_{r_{1}^{(k - 1)} - r_{1}^{(k)}}
(q)_{r_{1}^{(k)}}}$$ $$\sum_{r_{2}^{(1)}\geq \ldots \geq r_{2}^{(k)}
\geq 0}
\frac{q^{{r_{2}^{(1)}}^{2} + \ldots + {r_{2}^{(k)}}^{2} -
r_{2}^{(1)} r_{1}^{(1)} - \ldots - r_{2}^{(k)}r_{1}^{(k)}+
\sum_{t=1}^{k} r_{2}^{(t)}\delta_{2,j_{t}}  }}{(q)_{r_{2}^{(1)}
- r_{2}^{(2)}} \ldots (q)_{r_{2}^{(k - 1)}
- r_{2}^{(k)}} (q)_{r_{2}^{(k)}}}$$
$$\cdots\cdots\cdots$$
$$\sum_{r_{n}^{(1)}\geq \ldots \geq
r_{n}^{(k)} \geq 0}\;
\frac{q^{{r_{n}^{(1)}}^{2} + \ldots + {r_{n}^{(k)}}^{2} -
r_{n}^{(1)} r_{n - 1}^{(1)} - \ldots - r_{n}^{(k)}r_{n - 1}^{(k)}+
\sum_{t=1}^{k} r_{n}^{(t)}\delta_{n,j_{t}}
}}{(q)_{r_{n}^{(1)}
- r_{n}^{(2)}} \ldots (q)_{r_{n}^{(k - 1)}
- r_{n}^{(k)}} (q)_{r_{n}^{(k)}}}$$
(cf. (\ref{4.1}) where $j_{t}$ is introduced).

   Just like in the level one case, we can rewrite this formula in a
more compact matrix form. Namely, set $p_{i}^{(s)} := r_{i}^{(s)}  -
r_{i}^{(s+ 1)},$ $1 \leq s \leq k-1$ and $p_{i}^{(k)} :=
r_{i}^{(k)}$ (note that $p_{i}^{(s)}$ is exactly the number of
quasi-particles
of color $i$ and charge $s).$ Since $\hat{\Lambda} =  k_{0}\hat{\Lambda}_{0} +
k_{j}\hat{\Lambda}_{j}$ for some $j,\; 1\leq j \leq n,$ we have
$j_{t} = 0$ for $0 < t \leq k_{0}$ and $j_{t} = 1$ for $k_{0} < t \leq
k.$ A straightforward calculation now shows  that the above expession
can be rewritten as follows:
\be \mbox{\rm Tr}\,q^{D} \left| \begin{array}[t]{l} \\
W(k_{0}\hat{\Lambda}_{0} + k_{j}\hat{\Lambda}_{j}) \end{array} \right.
=\label{4.27}\ee $$= \sum_{\begin{array}{c}{\scriptstyle
p_{1}^{(1)},\ldots , p_{1}^{(k)} \geq 0}\\{\scriptstyle \ldots \ldots
\ldots\ldots}\\{\scriptstyle p_{n}^{(1)},\ldots
, p_{n}^{(k)} \geq 0 }\end{array}}
\frac{q^{\frac{1}{2}\sum_{l,m =
1, \ldots, n}^{s,t = 1,\ldots , k}\;A_{lm} B^{st}
p_{l}^{(s)}p_{m}^{(t)}}}{\prod_{i=1}^{n} \prod_{s=1}^{k}
(q)_{p_{i}^{(s)}} }\;q^{\tilde{p}_{j}}$$ where $$\tilde{p}_{j} :=
p_{j}^{(k_{0} +1)} + 2p_{j}^{(k_{0} + 2)} + \ldots + \
k_{j}p_{j}^{(k)},$$ $(A_{lm})_{l,m = 1}^{n}$ is the Cartan matrix of
$\g = sl(n + 1, \C)$ and $B^{st} := \mbox{min}\{s,t\},$ $1\leq s,t
\leq k.$

   In the case of the vacuum module $(\hat{\Lambda} =
k\hat{\Lambda}_{0}),$ this is the Feigin-Stoyanovsky character formula
announced in  [FS].

   In terms of a generating function with formal variables
$y_{1}^{(1)},\ldots y_{n}^{(k)}$ (respectively, $y_{1}, \ldots ,
y_{n}$)which encode the color-charge-type (resp., the color-type) of
the basis, one has
\be \mbox{\rm ch}\,
W(k_{0}\hat{\Lambda}_{0} + k_{j}\hat{\Lambda}_{j} )=\label{4.28}\ee
$$= \sum_{\begin{array}{c}{\scriptstyle p_{1}^{(1)},\ldots ,
p_{1}^{(k)} \geq 0}\\{\scriptstyle \ldots \ldots
\ldots\ldots}\\{\scriptstyle p_{n}^{(1)},\ldots
, p_{n}^{(k)} \geq 0 }\end{array}}
\frac{q^{\frac{1}{2}\sum_{l,m =
1, \ldots, n}^{s,t = 1,\ldots , k}\;A_{lm} B^{st}
p_{l}^{(s)}p_{m}^{(t)}}}{\prod_{i=1}^{n} \prod_{s=1}^{k} (q)_{p_{i}^{(s)}}
}\;q^{\tilde{p}_{j}}\prod_{i=1}^{n}\prod_{s=1}^{k}
(y_{i}^{(s)})^{p_{i}^{(s)}} = $$
$$= \sum_{\begin{array}{c}{\scriptstyle p_{1}^{(1)},\ldots
, p_{1}^{(k)} \geq 0}\\{\scriptstyle \ldots \ldots
\ldots\ldots}\\{\scriptstyle p_{n}^{(1)},\ldots
, p_{n}^{(k)} \geq 0 }\end{array}}
\frac{q^{\frac{1}{2}\sum_{l,m =
1, \ldots, n}^{s,t = 1,\ldots , k}\;A_{lm} B^{st}
p_{l}^{(s)}p_{m}^{(t)}}}{\prod_{i=1}^{n} \prod_{s=1}^{k} (q)_{p_{i}^{(s)}}
}\;q^{\tilde{p}_{j}}\prod_{i=1}^{n}
y_{i}^{\sum_{s=1}^{k} s p_{i}^{(s)}}.$$
The coefficient of
$$(y_{1}^{(1)})^{p_{1}^{(1)}} \cdots (y_{1}^{(k)})^{p_{1}^{(k)}}\cdots\cdots
(y_{n}^{(1)})^{p_{n}^{(1)}} \cdots
(y_{n}^{(k)})^{p_{n}^{(k)}}$$
in the first expression  gives the $q$-character of the subspace generated
by quasi-particle  monomials with exactly $p_{i}^{(s)}$
quasi-particles of color $i$
and charge $s$ (hence the color-charge-type of such monomials is
$$(\underbrace{1,\ldots ,1}_{p_{n}^{(1)}},\ldots ,
\underbrace{k,\ldots ,k}_{p_{n}^{(k)}};\ldots ; \underbrace{1,\ldots
,1}_{p_{1}^{(1)}},\ldots , \underbrace{k,\ldots ,k}_{p_{1}^{(k)}})$$
and the total charge is $\sum_{i=1}^{n}\sum_{s=1}^{k} s p_{i}^{(s)}).$
The coefficient of $y_{1}^{r_{1}}\cdots y_{n}^{r_{n}}$ in the second
expression gives the $q$-character of the weight subspace $W_{\Lambda
+
\sum_{i=1}^{n} r_{i}\alpha_{i}}(\hat{\Lambda}).$

\section {Appendix}

$\begin{array}{|c|c|c|} \hline \mbox{ \em color-} & \mbox{ \em energy}
&
\mbox{ \em basis}
\\ \mbox{\em -type} & & \\ \hline\hline   (1;2) & 3 & (1_{\alpha_{2}}
-3_{\alpha_{1}}-1_{\alpha_{1}}) \\ \cline{2-3} & 4 &
(1_{\alpha_{2}}-4_{\alpha_{1}}-1_{\alpha_{1}}), (0_{\alpha_{2}}-
3_{\alpha_{1}}-1_{\alpha_{1}})\\ \cline{2-3} & 5 &
(1_{\alpha_{2}}-5_{\alpha_{1}}-1_{\alpha_{1}}),
(1_{\alpha_{2}}-4_{\alpha_{1}}-2_{\alpha_{1}}),
(0_{\alpha_{2}}-4_{\alpha_{1}}-1_{\alpha_{1}}),\\ & &
(1_{\alpha_{2}}-3_{\alpha_{1}}-1_{\alpha_{1}}) \\ \hline (2;2) & 4 &
(-1_{\alpha_{2}}1_{\alpha_{2}}-3_{\alpha_{1}}-1_{\alpha_{1}}) \\
\cline{2-3} & 5 &
(-2_{\alpha_{2}}1_{\alpha_{2}}-3_{\alpha_{1}}-1_{\alpha_{1}}),
(-1_{\alpha_{2}} 1_{\alpha_{2}} -4_{\alpha_{1}} -1_{\alpha_{1}})
\\ \cline{2-3} & 6 &  (-3_{\alpha_{2}}1_{\alpha_{2}}-3_{\alpha_{1}}
-1_{\alpha_{1}}),
(-2_{\alpha_{2}}1_{\alpha_{2}}-4_{\alpha_{1}}-1_{\alpha_{1}}) ,
(-2_{\alpha_{2}}0_{\alpha_{2}}-3_{\alpha_{1}} -1_{\alpha_{1}}), \\ & &
(-1_{\alpha_{2}}1_{\alpha_{2}} -5_{\alpha_{1}} -1_{\alpha_{1}}),
(-1_{\alpha_{2}}1_{\alpha_{2}}-4_{\alpha_{1}}-2_{\alpha_{1}}) \\
\hline \end{array}$ \begin {center} Table 1 \end{center}\vspace{5mm}

\noindent$\begin{array}{|c|c|c|c|} \hline  \mbox{ \em color-} &  \mbox{ \em
energy} & \mbox{\em color-} & \mbox{\em basis}\\ \mbox{\em -type} & &
\mbox{\em -charge-} & \\ & & \mbox{\em -type} &\\
\hline \hline (1;2) & 2 &(1;2)& (0_{\alpha_{2}}-2_{2\alpha_{1}})\\
\cline{2-4} & 3
& (1;1,1)& (1_{\alpha_{2}}-3_{\alpha_{1}}-1_{\alpha_{1}})\\
\cline{3-4}& & (1;2)&
(0_{\alpha_{2}}-3_{2\alpha_{1}}),(-1_{\alpha_{2}}-2_{2\alpha_{1}})\\
\cline{2-4} & 4 &(1;1,1) &  (1_{\alpha_{2}}-4_{\alpha_{1}}-1_{\alpha_{1}}),
(0_{\alpha_{2}}-3_{\alpha_{1}}-1_{\alpha_{1}})\\\cline{3-4} & & (1;2)
& (0_{\alpha_{2}}-4_{2\alpha_{1}}), (-1_{\alpha_{2}}
-3_{2\alpha_{1}}), (-2_{\alpha_{2}}-2_{2\alpha_{1}}) \\ \cline{2-4} &
5 & (1;1,1) & (1_{\alpha_{2}} -4_{\alpha_{1}} -2_{\alpha_{1}}),
(1_{\alpha_{2}} -5_{\alpha_{1}} -1_{\alpha_{1}}),\\ & &
&(0_{\alpha_{2}} -4_{\alpha_{1}}-1_{\alpha_{1}}),
(-1_{\alpha_{2}}-3_{\alpha_{1}}-1_{\alpha_{1}})\\\cline{3-4}& & (1;2)&
(0_{\alpha_{2}} -5_{2\alpha_{1}}), (-1_{\alpha_{2}} -4_{2\alpha_{1}}),
(-2_{\alpha_{2}} -3_{2\alpha_{1}}),\\ & & & (-3_{\alpha_{2}}
-2_{2\alpha_{1}})\\
\hline (2;2) & 2 & (2;2) & (0_{2\alpha_{2}}-2_{2\alpha_{1}})\\\cline{2-4}& 3
& (2;2) & (0_{2\alpha_{2}} -3_{2\alpha_{1}}), (-1_{2\alpha_{2}}
-2_{2\alpha_{1}})\\ \cline{2-4} & 4 & (1,1;1,1) & (-1_{\alpha_{2}}
1_{\alpha_{2}}-3_{\alpha_{1}}-1_{\alpha_{1}})\\ \cline{3-4} & &
(2;1,1) & (0_{2\alpha_{2}}-3_{\alpha_{1}}-1_{\alpha_{1}})\\
\cline{3-4} & & (1,1;2) & (-2_{\alpha_{2}} 0_{\alpha_{2}}
-2_{2\alpha_{1}})\\ \cline{3-4} & & (2;2) & (0_{2\alpha_{2}}
-4_{2\alpha_{1}}), (-1_{2\alpha_{2}} -3_{2\alpha_{1}}),
(-2_{2\alpha_{2}} -2_{2\alpha_{1}})\\ \cline{2-4} & 5 & (1,1;1,1) &
(-1_{\alpha_{2}} 1_{\alpha_{2}} -4_{\alpha_{1}}
-1_{\alpha_{1}}),(-2_{\alpha_{2}} 1_{\alpha_{2}}
-3_{\alpha_{1}}-1_{\alpha_{1}})\\ \cline{3-4} & & (2;1,1) &
(0_{2\alpha_{2}} -4_{\alpha_{1}}-1_{\alpha_{1}}), (-1_{2\alpha_{2}}
-3_{\alpha_{1}} -1_{\alpha_{1}}) \\ \cline{3-4} & & (1,1;2) &
(-2_{\alpha_{2}} 0_{\alpha_{2}} -3_{2\alpha_{1}}),(-3_{\alpha_{2}}
0_{\alpha_{2}} -2_{2\alpha_{1}}) \\ \cline{3-4} & & (2;2) &
(0_{2\alpha_{2}} -5_{2\alpha_{1}}), (-1_{2\alpha_{2}}
-4_{2\alpha_{1}}), (-2_{2\alpha_{2}} -3_{2\alpha_{1}}),\\ & & &
(-3_{2\alpha_{2}} -2_{2\alpha_{1}})\\ \hline \end{array}$
\begin {center} Table 2 \end{center}


\vspace{10mm}




\begin{thebibliography}{AAAAAA}
\bibitem[A]{1} Andrews, G. E.: The theory of partitions.
Addison-Wesley 1976
\bibitem[B]{17} Borcherds, R.E.: Vertex algebras, Kac-Moody algebras
and the Monster. Proc. Natl. Acad. Sci. USA {\bf 83}, 3068-3071 (1986)
\bibitem[Ber]{35} Berkovich, A.:  Fermionic counting of RSOS-states and
Virasoro character formulas for the unitary minimal series {\cal
M}($\nu , \nu +1).$ Exact results. Nucl. Phys. {\bf B431}, 315
(1994); hep-th/9403073
\bibitem[BLS1]{} Bouwknegt, P., Ludwig, A.  and Schoutens, K.: Spinon bases,
Yangian symmetry and fermionic representations of Virasoro characters
in conformal field theory. Phys. Lett.  {\bf 338B}, 448  (1994);
hep-th/9406020
\bibitem[BLS2]{} Bouwknegt, P.,  Ludwig, A. and Schoutens, K.: Spinon
basis for higher level $SU(2)$ WZW models. Preprint USC-94/20;
hep-th/9412108
\bibitem[BLS3]{} Bouwknegt, P., Ludwig, A.  and Schoutens, K.: Affine and
Yangian symmetries in $SU(2)_{1}$ conformal field theory. Preprint USC-94/21;
hep-th/9412199
\bibitem[BM]{} Berkovich, A.  and McCoy, B.: Continued fractions and
fermionic representations for characters of $M(p,p')$ minimal models.
Preprint BONN-TH-94-28; hep-th/9412030
\bibitem[BPS]{} Bernard, D.,  Pasquier, V.  and Serban,D.: Spinons in
Conformal Field Theory. Nucl. Phys.  {\bf B428}, 612  (1994);
hep-th/9404050.
\bibitem[BPZ]{36} Belavin, A.,  Polyakov, A. and  Zamolodchikov, A.: Infinite
conformal symmetry in two-dimensional quantum field theory. Nucl.
Phys.  {\bf B241}, 333-380  (1984)
\bibitem[C]{} Capparelli, S.: On some representations of twisted affine
Lie algebras and combinatorial identities. J. Algebra {\bf 154},
335-355 (1993)
\bibitem[DKKMM1]{} Dasmahapatra, S., Kedem, R.,  Klassen, T.R., McCoy,
B.  and Melzer, E.:  Quasi-particles, conformal field theory and q
series. Int. J. Mod. Phys. {\bf B7},3617 (1993);
hep-th/9303013
\bibitem[DKKMM2]{} Dasmahapatra, S.,  Kedem, R., Klassen, T.R.,  McCoy, B.
and Melzer, E.: Virasoro characters from Bethe  equations for the
critical ferromagnetic three-state Potts model. J. Stat. Phys.
{\bf  74},239  (1994); hep-th/9304150
\bibitem[DL]{4} Dong, C.  and Lepowsky, J.:  Generalized
Vertex Algebras and Relative Vertex Operators. Progress in Math. {\bf
112} Birkhauser 1993
\bibitem[FF]{} Feigin, B.  and Frenkel, E.:  Coinvariants of nilpotent
subalgebras of the Virasoro algebra and partition identities. Adv.
Sov. Math. {\bf 16}, 139-148 (1993)
\bibitem[FHL]{5} Frenkel, I.,  Huang, Y.-Z. and  Lepowsky, J.:  On
axiomatic approaches to vertex operator algebras and modules.
preprint  1989; Memoirs Amer. Math. Society {\bf 104} (1993)
\bibitem[FK]{18} Frenkel, I.  and Kac, V.: Basic representations of
affine Lie algebras and dual resonance models. Invent. Math.
{\bf 62}, 23-66  (1980)
\bibitem[FL]{} Feingold, A. and Lepowsky, J.: The Weyl-Kac character
formula and power series identities. Adv. Math. {\bf 29}, 271-309
(1978)
\bibitem[FLM]{6} Frenkel, I.,  Lepowsky, J. and Meurman, A.:  Vertex
Operator Algebras and the Monster.  Pure and Appl. Math. {\bf 134}.
Academic Press 1988
\bibitem[FNO]{7} Feigin, B., Nakanishi, T.  and Ooguri, H.:
The annihilating ideals of minimal models. Int. J. Mod. Phys. {\bf
A7}, 217-238 (1992)
\bibitem[FQ]{} Foda, O.  and Quano, Y.:  Virasoro character identities
from the Andrews-Bailey construction. Preprint hep-th/9408086
\bibitem[FQS]{100} Friedan, D., Qiu, Z.  and Shenker, S.:  Conformal
invariance,
unitarity and two-dimensional critical exponents. In: J. Lepowsky et
al. (eds.) Vertex
Operators in Mathematics and Physics. MSRI
publication no.3 1985, pp. 419-450; Phys. Rev. Lett   {\bf
52}, 1575  (1984)
\bibitem[FS]{8} Feigin, B.  and Stoyanovsky, A.:
Quasi-particles models for the representations of Lie algebras and
geometry of flag manifold. Preprint RIMS-942  1993; hep-th/9308079;
cf. also the short version: Functional models for representations of
current algebras and semi-infinite Schubert cells. Func. Anal.
Appl. {\bf 28} No. 1, 15  (1994)
\bibitem[FW]{} Foda, O.  and Warnaar, S.: A bijection which implies
Melzer's polynomial identities: the $\chi_{1,1}^{(p,p+1)}$ case.
Preprint hep-th/9501088
\bibitem[G]{9} Gepner, D.:  New conformal field theories associated
with Lie algebras and their partition functions. Nucl. Phys.
{\bf B290}, 10-24  (1987)
\bibitem[GeII]{}Georgiev,  G.: Combinatorial  constructions of
modules for infinite dimensional Lie algebras, II. Parafermionic
space.  To appear
\bibitem[GeIII]{10}Georgiev, G.: Fermionic  characters for Virasoro
algebra modules. In preparation
\bibitem[GKO]{53} Goddard, P.,  Kent, A.  and Olive, D.: Unitary
representations of the Virasoro and super-Virasoro algebras. Commun.
Math. Phys.  {\bf 103}, 105-119  (1986)
\bibitem[Go]{52} Gordon, B.:  A combinatorial generalization of the
Rogers-Ramanujan
identities. Amer. J. Math.  {\bf 83}, 393-399  (1961)
\bibitem[H]{} Haldane, F.D.M.: ''Fractional statistics'' in arbitrary
dimensions: a generalization of the Pauli principle. Phys. Rev.
Lett. {\bf 67}, 937-940  (1991)
\bibitem[Hu]{} Husu,  C.: Extensions of the Jacobi identity for vertex
operators,
and standard $A_{1}^{(1)}$-modules. Memoirs Amer. Math. Society
{\bf 106} (1993)
\bibitem[K]{10} Kac, V.:  Infinite-dimensional Lie
algebras.  Cambridge University Press 1990
\bibitem[Kir]{} Kirillov, A.N.:  Dilogarithm identities. Preprint
hep-th/9408113
\bibitem[KKMM1]{3} Kedem, R.,  Klassen, T.,
McCoy, B. and Melzer, E.: Fermionic quasi-particle representations for
characters of $\frac{(G^{(1)})_{1} \times
(G^{(1)})_{1}}{(G^{(1)})_{2}}.$ Phys. Lett.  {\bf B304}, 263-270
(1993); hep-th/9211102
\bibitem[KKMM2]{} Kedem, R.,  Klassen, T.,
McCoy, B. and  Melzer, E.:  Fermionic sum representations for conformal field
theory characters.  Phys. Lett. {\bf B307}, 68-76 (1993);
hep-th/9301046
\bibitem[KM]{} Kedem, R. and  McCoy, B.: Construction of modular branching
functions from Bethe's equation in the 3-state  Potts chain. J.
Stat. Phys.  {\bf 71}, 865  (1993); hep-th/9210129
\bibitem[KMM]{} Kedem, R.,  McCoy, B. and  Melzer, E.:  The Sums of
Rogers, Schur
and Ramanujan and the Bose-Fermi correspondence in 1+1-dimensional
quantum field theory. Preprint hep-th/9304056
\bibitem[KNS]{11} Kuniba,  A., Nakanishi, T.  and Suzuki, J.:
Characters of conformal field theories from thermodynamic Bethe
ansatz Mod. Phys. Lett.  {\bf A8}, 1649-1660  (1993);
hep-th/9301018
\bibitem[LP1]{} Lepowsky, J.  and Primc, M.:  Standard modules for type
one affine algebras. Lecture Notes in Math.  {\bf 1052}, 194-251
(1984)
\bibitem[LP2]{13} Lepowsky, J.  and Primc, M.:  Structure of the
standard modules for the affine algebra $A_{1}^{(1)}.$ Contemp.
Math.  {\bf 46} (1985)
\bibitem[LW1]{56} Lepowsky, J.  and  Wilson, R.:  A new family of algebras
underlying the Rogers-Ramanujan identities and generalizations.
Proc. Natl. Acad. Sci. USA {\bf 78}, 7245-7248  (1981)
\bibitem[LW2]{57} Lepowsky, J.  and Wilson, R.:  The structure of standard
modules, I: Universal algebras and Rogers-Ramanujan identities.
Invent. Math.  {\bf 77}, 199-290  (1984)
\bibitem[LW3]{58} Lepowsky, J.  and Wilson, R.:  The structure of standard
modules, II: The case $A_{1}^{(1)},$ principal gradation.
Invent. Math.  {\bf 79}, 417-442  (1985)
\bibitem[M1]{34} Melzer, E.:  Fermionic character sums and the corner
transfer matrix. Int. J. Mod. Phys.  {\bf A9}, 1115-1136 (1994);
hep-th/9305114.
\bibitem[M2]{} Melzer, E.:  The many faces of a character. Lett.
Math. Phys.  {\bf 31}, 233-246  (1994); hep-th/9312143.
\bibitem[Ma]{} Mandia, M.:  Structure of the level one standard modules
for the affine Lie algebras $B_{l}^{(1)},\; F_{4}^{(1)}$ and
$G_{2}^{(1)}. $ Memoirs Amer. Math. Society
{\bf 362}  (1987)
\bibitem[Mi1]{} Misra, K.:  Realization of the level two standard
$sl(2k+1,\C)^{~}$-modules. Trans. Amer. Math. Soc.  {\bf 316}, 295-309
(1989)
\bibitem[Mi2]{} Misra, K.:  Realization of the level one  standard
$\tilde{C}_{2}$-modules.   Trans. Amer. Math. Soc.  {\bf 321}, 483-504
(1990)
\bibitem[Mi3]{}Misra, K.:  Level  one  standard modules for affine
symplectic Lie algebras.  Math. Ann.  {\bf 287}, 287-302 (1990)
\bibitem[Mi4]{}Misra, K.:  Level  two  standard
$\tilde{A}_{n}$-modules. J. Algebra  {\bf 137}, 56-76  (1991)
\bibitem[MP1]{} Meurman, A.  and Primc, M.:  Annihilating ideals of
standard modules of $sl(2,\C)\tilde{}$ and combinatorial identities. Adv.
Math.  {\bf 64}, 177-240  (1987)
\bibitem[MP2]{} Meurman, A.  and Primc, M.:  Annihilating fields  of
standard modules of $sl(2,\C)\tilde{}$ and combinatorial identities.
Preprint 1994
\bibitem[NRT]{} Nahm, W., Recknagel,  A. and  Terhoeven, M.:  Dilogarithm
identities in conformal field theory. Mod. Phys. Lett.  {\bf A8},
1835-1848 (1993)
\bibitem[P1]{} Primc, M.:  Standard representations of $A_{n}^{(1)}.$
In: Kac, V. (ed.)  Infinite-dimensional Lie algebras and groups,
Advanced Series in Math. Physics {\bf 7} 1989, pp. 273-282.
\bibitem[P2]{55} Primc, M.:  Vertex operator construction of standard
modules for $A_{n}^{(1)}.$ Pacific J. Math. {\bf 162}  No.1, 143-187
(1994)
\bibitem[S]{19} Segal, G.:  Unitary representations of some
infinite-dimensional groups. Commun. Math. Phys.  {\bf 80}, 301-342
(1981)
\bibitem[T]{15} Terhoeven, M.: Lift of dilogarithm to partition
identities. Preprint BONN-HE-92-36  1992; hep-th/9211120
\bibitem[WP]{} Warnaar, S.  and Pearce, P.:  A-D-E polynomial and
Rogers-Ramanujan identities. Preprint hep-th/9411009
\bibitem[ZF1]{16} Zamolodchikov, A.B.  and Fateev, V.A.:
Non-local (parafermion) currents in two-dimensional quantum field
theory and self-dual critical points in $\Z_{n}$- symmetric classical
systems. Sov. Phys. JETP  {\bf 62}, 215  (1985)
\bibitem[ZF2]{59} Zamolodchikov, A.B.  and Fateev, V.A.:  Disorder fields
in two-dimensional conformal quantum field theory and  $N=2$ extended
supersymmetry. Sov. Phys. JETP  {\bf 63}, 913-919  (1986)

\end{thebibliography}
\end{document}